\documentclass[10]{article}

\usepackage{graphicx}
\usepackage{psfrag}
\usepackage{amssymb}
\usepackage{amsmath}

\newcommand{\rf}[1]{(\ref{#1})}
\newcommand{\beq}{\begin{equation}}

\newcommand{\eeq}{\end{equation}}
\newcommand{\bea}{\begin{eqnarray}}
\newcommand{\eea}{\end{eqnarray}}

\newcommand{\e}{\mbox{e}}
\renewcommand{\d}{\mbox{d}}

\newcommand{\lam}{\lambda}
\newcommand{\Lam}{\Lambda}

\newcommand{\al}{\alpha}

\renewcommand{\th}{\theta}
\newcommand{\ep}{\varepsilon}
\newcommand{\eps}{\epsilon}
\newcommand{\om}{\omega}

\newcommand{\del}{\delta}
\newcommand{\Del}{\Delta}
\newcommand{\sg}{\sigma}
\newcommand{\kp}{\kappa}

\newcommand{\oh}{\frac{1}{2}}
\newcommand{\oq}{\frac{1}{4}}

\newcommand{\ra}{\rangle}
\newcommand{\la}{\langle}

\newcommand{\mi}{\!-\!}
\newcommand{\equ}{\!=\!}
\newcommand{\plu}{\!+\!}

\newcommand{\cD}{{\cal D}}

\newcommand{\cN}{{\cal N}}

\begin{document}
%
%
%

\begin{center}
\vspace{24pt}
{ \large \bf Lattice Quantum Gravity: EDT and CDT}\footnote{This is a contribution to the {\it Handbook of Quantum Gravity} which 
will be published in  2023. It will appear as a chapter in the section of the handbook denoted {\it Causal Dynamical 
triangulations}.} 

\vspace{24pt}

{\sl J. Ambjorn}$\,^{a,b}$

\vspace{10pt}

{\small

$^a$~The Niels Bohr Institute, Copenhagen University\\
Blegdamsvej 17, DK-2100 Copenhagen \O , Denmark.\\
email: ambjorn@nbi.dk

\vspace{10pt}

$^b$~Institute for Mathematics, Astrophysics and Particle Physics
(IMAPP)\\ Radboud University Nijmegen, Heyendaalseweg 135, \\
6525 AJ  Nijmegen, The Netherlands\\

}

\end{center}

\vspace{24pt}

\begin{center}
{\bf Abstract}
\end{center}

\vspace{11pt} 

\noindent 
This article is  an overview of the use of so-called Euclidean Dynamical Triangulations  (EDT)
and Causal Dynamical Triangulations (CDT)
as  lattice regularizations of quantum gravity. The lattice regularizations have been very successful in the case of 
two-dimensional quantum gravity, where the lattice theories indeed provide  regularizations of 
continuum well defined quantum gravity theories. In four-dimensional spacetime the Einstein-Hilbert action leads to a 
theory of gravity which is not renormalizable as a perturbative quantum theory around flat spacetime.   
It is discussed how lattice gravity in the form of EDT or CDT can be used 
to search for a non-perturbative UV fixed point of the lattice renormalization group in the spirit of asymptotic safety. 
In this way it might be possible to define a  quantum theory of gravity also at length
scales smaller than the Planck length.

\newpage

\section{Introduction}

So far there is no universally accepted quantum theory of four-dimensional gravity. The classical theory of general relativity
is not perturbative renormalizable. Therefore, if we think about four-dimensional quantum gravity as small quantum fluctuations around 
some classical geometry that solves Einstein's equations, it  
only makes sense as an effective quantum field
theory  up to some energy or down to some length-scale, determined by the coupling constants entering in the classical theory
(this is discussed in detail in the Section ``Effective Quantum Gravity" in the Handbook).
Unfortunately, we have presently no experiments which can guide us if we want to approach or go beyond this scale,
which is the Planck energy or the Planck length\footnote{If $G$ denotes Newton's gravitational constant, $c$ the velocity of light
and $\hbar$ the Planck constant, the Planck energy is $E_p = \sqrt{\hbar c^5/G}$ and the Planck length is 
$\ell_p = \sqrt{\hbar G/c^3}$.}. 
There are examples of other quantum field theories which, when viewed 
at low energies, appeared to be non-renormalizable, like the theory of weak interactions and the theory of strong interactions. 
In the case of the weak interactions, the four-fermion interaction, originally suggested to explain the weak interactions, 
is non-renormalizable. However, we now know that at high energies it is resolved into renormalizable interactions mediated
by $W$ and $Z$ particles. Similarly, the non-renormalizable non-linear sigma model, which was used to described the low energy 
$\pi$-$\pi$ interaction related to the strong interactions, is a low-energy effective action of an underlying renormalizable 
quantum field theory of quarks and gluons. From these examples it is tempting to conjecture that the same could happen 
to quantum gravity, and that the non-renormalizability of gravity would be resolved at larger energy by new degrees of 
freedom that we have not yet observed. It could be the case, but gravity still looks different from these two examples.
In the case of the weak interactions one  was led to the four-fermion interaction and not to
the renormalizable version of the weak interactions simply because the  $W$ and $Z$ particles were so heavy 
that they had not yet been observed. In the case of the strong interactions one was led to a $\pi$-$\pi$ interaction because the 
quark and gluons were not observed, not because they were heavy, but because of quark and gluon confinement. In both 
cases the starting points were really  (effective) quantum theories, the classical aspects of the theories playing minor roles. In 
gravity the situation is different. We have a classical theory, which seemingly works very well, and this theory even has
long-range massless classical excitations propagating with the velocity of light in a classical background 
geometry, the now famous gravitational waves. It does not seem too promising to try to explain this as a limit 
of a renormalizable quantum theory constructed from heavy, yet to be observed, fundamental particles, or from
``confined'' light particles. 

String theory is one attempt to provide a quantum theory of gravity.
More precisely, closed string theory contains  massless 
spin-two excitations, which can be interpreted as  quantum gravity particles and  the underlying stringy nature of the 
theory solves the UV problems associated with quantizing the Einstein-Hilbert action of classical gravity. 
The original hope was that the (super)string
theory would provide us with an explanation of all the particles we actually observe in nature, and at the same time it  predicted
the existence of particles we have not yet observed. Unfortunately,  no clear picture related to the world we observe 
has yet emerged from string theory, which, since the ambitious start in the 1980s as a Theory of Everything, has developed 
in many directions. The directions still related to gravity will be described in the Section of the Handbook dedicated to 
string theory.  

Loop quantum gravity is another attempt to circumvent the problem associated with a ``naive'' quantization of gravity based 
on the Einstein-Hilbert action. It deals with the UV problems of the naive approach by  postulating a  new quantization
procedure which leads to a Hilbert space quite different from the standard Fock space of particle physics. This procedure 
defines in principle the physics at the Planck scale, but it becomes difficult to relate the theory to classical gravity as 
we observe it today. Like string theory it has branched out in a number of different directions, again described 
in the Handbook in the Section about loop quantum gravity.

Lattice quantum gravity, as described in this Section, is closely related to a field theoretical approach to quantum 
gravity using what is called {\it asymptotic safety}. A special Section in the Handbook is dedicated to the use 
of asymptotic safety when the quantizing gravity. The hypothesis is that one can use ordinary quantum field theory
to quantize gravity and that the UV limit of the theory corresponds to a {\it non-perturbative fixed point}. Around this fixed
point one cannot apply ordinary perturbation theory, by expanding in a power series of coupling constants appearing in 
the classical  low-energy Lagrangian. Nevertheless it is postulated that there exists a continuum renormalization group flow
of the effective action that will reach the UV fixed point by  adjusting only a finite number of suitable coupling constants.
In this sense the non-perturbative fixed point is similar to a (Gaussian) UV fixed point  of a 
renormalizable quantum field theory. It is in this context that lattice quantum gravity becomes interesting for a number of reasons.

From a Wilsonian point of view lattice field theory is well suited to studying  fixed points and renormalization group flows, as well as
non-perturbative aspects of the corresponding quantum field theories. The lattice will provide a UV regularization of the 
quantum field theory in question. Such a UV regularization is  usually needed as starting point for defining an
interacting  quantum field theory. In order to define the corresponding continuum quantum field theory one will in general 
need to take the UV cut-off, i.e.\ the lattice spacing, to zero relative to some continuum length scale characterizing the continuum
theory. If the theory contains massive particles one can use the inverse mass of such a particle as the length scale (in units
where $c = \hbar =1$). Keeping such a physical length scale fixed while taking the lattice spacing to zero implies that this 
length scale measured in lattice units will diverge. This is  most simply realized in lattice field theories  if  a correlator
of one of the lattice fields for a generic choice of the lattice coupling constants $g_i$ of the theory is decaying exponentially with the 
distance between the lattice points, in this way defining a correlation length $\xi(g_i)$. A second- (or higher-) order phase transition 
of the lattice field theory is often characterized by a divergent correlation length. Thus, when trying to find continuum limits 
of the lattice field theory, it is natural to look for regions in the lattice coupling space associated 
with second- or higher-order 
phase transitions. 
One of the assumptions in the Wilsonian approach, if one consider lattice field Hamiltonians with arbitrary local interactions, i.e.\
in principle an infinite-dimensional coupling constant space, is that  the critical surface of higher-order phase transitions has 
a finite co-dimension. In this case one only has to fine-tune a finite number of coupling constants to reach the critical surface 
where the lattice correlation length is infinite. The way in which one approaches the critical surface will define the continuum 
physical parameters of the corresponding continuum quantum field theory (like the masses of the particles corresponding 
to lattice field correlators), and the nature and the number of lattice coupling constants which  need to be fine-tuned
to reach the critical surface will depend  on the so-called  {\it fixed points} of the lattice renormalization group. 
These fixed points are located on the critical surface, and in the Wilsonian 
picture each  can be used to define  a continuum quantum field theory.

We want to use this lattice  Wilsonian framework to investigate whether we can define a continuum limit of 
theories we can denote ``lattice gravity'' theories,  more precisely, the lattice gravity theories based on Euclidean
Dynamical Triangulations (EDT) or Causal Dynamical Triangulations (CDT). We  have a space of (dimensionless) lattice 
coupling constants associated with the theory and  want to locate regions in this space where there are second- (or higher-) order 
phase transitions. When such regions are localized, we want to understand whether one can approach these phase transition
regions such that one obtains a theory that can be viewed as the quantum theory of gravity. It is of particular interest if 
the  phase transition surface can be associated with a UV fixed point, since in this case one might have defined
the quantum gravity theory at arbitrarily  short distances.

A number of interesting conceptual problems are associated with quantum gravity and the Wilsonian lattice renormalization group.
The central Wilsonian idea is that  a divergent lattice correlation length of some observable  makes it possible to forget the 
underlying lattice and that using a limiting procedure makes it possible to define a continuum quantum field theory.  It is also
the reason for the universality associated with the Wilsonian approach: the details of the local lattice structure as well as the details
of the interactions at lattice distances are often of no consequence for the continuum limit. Global symmetries of the 
interactions might be (and are) important as they can survive when a continuum limit 
is taken. However, when 
trying to apply this line of reasoning
to a lattice gravity theory we are faced with the very simple question: how does one define a correlation length in a theory
of quantum gravity? When implementing the quantum theory via a path integral, we are instructed to integrate over all geometries,
but it is the geometries which define the distances. In non-gravitational relativistic 
quantum field theory, correlators are functions of spacetime points, and 
the main reason we study these correlators is that their behavior as a function of the distances between these spacetime 
points tells us a lot about the underlying quantum theory. They are also the natural objects on which the renormalization group 
acts. Thus it is somewhat disturbing that it is unclear how to define such correlators in a theory of quantum gravity 
in a way that relates to {\it distances}. A common, and in general healthy, attitude in theoretical physics is to calculate whatever 
can be calculated and postpone annoying questions like how to define distances in a theory of quantum gravity. However,
one nice thing about a lattice theory of gravity is that one is forced to address such questions. Four-dimensional lattice gravity
cannot in any way be solved analytically, but one can perform computer simulations of the lattice theory. If one wants 
to measure anything but the simplest global observables in such computer simulations, one should better have a precise idea of how 
to define the observable to measure in a sensible way. In the rest of this Chapter we will discuss lattice quantum gravity from this 
Wilsonian point of view and how one can in principle use the lattice approach to test the asymptotic safety conjecture.
.

\section{A Wilsonian view on two-dimensional EDT and CDT}

\subsubsection*{The EDT partition functions}

Two-dimensional gravity is classically a trivial theory since there are no propagating gravitons in two-dimensional
spacetime. One reflection of this is that the curvature term in the Einstein-Hilbert action is a topological invariant in two dimensions. 
Thus, as long as one does not consider topology changes, and we will not do that,  the action just contains the cosmological
term, a term with no derivatives of the metric. If we consider spacetimes with Euclidean signature, we have 
the two-dimensional partition function
\beq\label{j1}
Z(G,\Lam,Z_i) = \int \cD [g] \; \e^{-S[g]}, 
\eeq
\beq\label{j2}
S[g] = - \frac{1}{2\pi G} \int d^2 \xi \sqrt{g} (R(\xi) - 2 \tilde{\Lam}) = -\frac{\chi}{G} + \Lam V[g] + \sum_{i=1}^nZ_i L_i[g].
\eeq
The path integral \rf{j1} is over all  {\it geometries} $[g_{ab}]$ (i.e.\ metrics $g_{ab}$ up to diffeomorphism equivalence) 
on a two-dimensional 
 manifold with $h$ handles and $n$ boundaries and with   
 Euler characteristic $\chi = 2-2h -n$. $\Lam$ is the cosmological 
 constant (divided by $\pi G$) and $Z_i$ are suitable boundary cosmological constants, 
 which are only introduced for 
 later convenience.  $V[g]$ denotes the two-dimensional volume of the 
 manifold, while $L_i[g]$ denotes the length
 of the $i$th boundary, all measured in the geometry $[g_{ab}]$. In the following we will ignore the topological term
 and only consider manifolds with the topology of a sphere with boundaries. 
 The partition function \rf{j1} can be written as 
 \bea\label{j3}
 Z(\Lam,Z_i) &=& \int_0^\infty dV \int_0^\infty \prod_{i=1}^n d L_i \; \e^{-\Lam V-L_i Z_i} 
 \int \cD_{V,L_i}  [g]  \\
 &=& \int_0^\infty dV \int_0^\infty \prod_{i=1}^n d L_i \; \e^{-\Lam V-L_i Z_i}  \cN(V,L_i). \label{j4}
 \eea
$[g]$ denotes a geometry and  
in \rf{j3} the functional integration is over all geometries which have 
spacetime volume $V$ and $n$ boundaries of 
 lengths $L_i$.  This integration is formally equal to the {\it number} of such geometries. 
 In other words, we can compute the partition
 function of two-dimensional quantum gravity if we can {\it count} the number of geometries with a given spacetime volume and 
 given lengths of the boundaries. Moreover, from this perspective the partition function $Z(\Lam,Z_i)$ can be viewed as 
 the generating function for the numbers $ \cN(V,L_i)$, with $e^{-\Lam}$ and $e^{-Z_i}$ playing the role of indeterminates in this generating function.
 Of course $ \cN(V,L_i)$ is formally infinite, reflecting the fact that the path integral $\int \cD_{V,L_i}  [g] $ needs a UV cut off to be 
 defined in the first place. The so-called (Euclidean) dynamical triangulations (EDT)
 provide a useful regularization\footnote{Historically, the  main interest in the EDT regulatrization was linked to the use
 as a regularization of the Polyakov path integral for the bosonic string in $D$-dimensional 
 spacetime\cite{adf,adfo,adf1,bd,djkp,kkm,bkkm}. This path integral  can be viewed as two-dimensional quantum gravity 
 coupled to $D$ bosonic fields $X_i$, constituting the $D$ coordinates of  the bosonic string. Unfortunately, the approach did not 
 work when implemented in the simplest way  for $D > 1$ as shown in \cite{ad}. However, for $D < 1$ it has been very successful, 
 and known as ``non-critical'' string theory, as will be mentioned below. 
 Pure two-dimensional quantum gravity, which we discuss here, corresponds in this context to $D=0$, and 
 was first introduced in \cite{david} and discussed in \cite{kkm}. There are interesting indications that the formalism can be 
 revived as a regularization of bosonic strings for $D>1$ by taking a new kind of scaling limit \cite{am3,am4,am5}.}. In its simplest version
 one approximates the integration over geometries by a summation over triangulations constructed from 
 equilateral triangles with link length $a$, where $a$  serves as a UV cut-off. To each such triangulation 
 one can associate a piecewise linear geometry by assuming the triangles are flat in the interior. The curvature of 
 the piecewise linear geometry is then naturally located at the vertices. Summing over such triangulations one 
 obtains an approximation to the continuum partition function \rf{j1} when one makes the identification 
 \beq\label{j5}
 V(T) = \frac{\sqrt{3}}{2} \,  N(T) a^2, \qquad L_i(T) = l_i (T) a,
 \eeq
 where $N(T)$ is the number of triangles and $l_i(T)$ the number of boundary links of the $i$th boundary in the triangulation $T$. 
 The lattice gravity partition function can be written as\footnote{It is assumed that a link on each boundary is marked, in order to 
 avoid symmetry factors appearing in the sum over triangulations.} 
 \beq\label{j5a}
 Z(\mu,\lam_i)= \sum_T \e^{-\mu N(T) - \sum_i \lam_i l_i(T)} = \sum_{N, l_i} \e^{-\mu N - \sum_i \lam_i l_i} \cN(N,l_i),
  \eeq
  which is the lattice version of \rf{j3} and \rf{j4}, the integration over geometries being replaced by the summation over 
  equilateral triangles with link length $a$, and with
  \beq\label{j5b}
  \mu = \Lam_0 a^2, \qquad  \lam_i = Z_i^{(0)} a.
  \eeq 
  We call $\Lam_0$ and $Z_i^{(0)}$ the bare, unrenormalized coupling constants for reasons that will become 
  clear below. 
  By counting the number
 of triangulations, $\cN(N,l_i)$, with the topology of a sphere with $n$ boundaries, and performing the sum and eventually taking the limit $a \to 0$, one 
 can then explicitly  find the partition function of two-dimensional quantum gravity.

 Let us  discuss  the Wilsonian aspect of the above procedure. From the Wilsonian point of view, the continuum limit should 
 not depend in a crucial way on precisely which class of triangulations one chooses. Similarly, one should be able not only to 
 use triangles, but also squares, pentagons etc.\ as  building blocks, all with link lengths $a$. One then looses the 
 unique piecewise geometry associated with a given graph $T$, but in the Wilsonian spirit one would still assume that 
 for very large graphs one can make an identification like in eq.\ \rf{j5}:
 \beq\label{j5c}
 V(T) \propto  N(T) a^2, \qquad L_i(T)  \propto l_i (T) a,
 \eeq
 where $N(T)$ denotes the number of polygons in the graph $T$ and $l_i$ the number of links of the $i$th boundary.
 This turns out to be true. For a particular 
 set of graphs, so-called bipartite graphs\footnote{In this context we define the bipartite graphs as surfaces constructed by gluing 
 together polygons with an even number of links, and where also the boundary loops consist of an even number of links.} 
 (again with the topology of a sphere with $n$ boundaries), 
 one can even find the corresponding generating function explicitly \cite{ajm,book},
 \beq\label{j6}
 Z(g,z_i) = \Big(\frac{1}{M_1(c^2,g)} \frac{d}{d c^2}\Big)^{n-3} \frac{1}{2c^2 M_1(g,c^2)} \prod_{i=1}^n \frac{c^2}{(z_i^2 - c^2)^{3/2}},
 \qquad n \geq 3.
 \eeq 
 In this expression we have assigned the indeterminate $g_k = g w_k$ to each $2k$-edged polygon which enters in the graph, 
 and an indeterminate $1/z_i$ to each link in the $i$th boundary.   The relative weights of the polygons are $w_k \geq 0$ and
 \beq\label{j7}
 M_1(c^2,g) =  \oint_C \frac{dz}{2\pi i} \; \frac{z V'(z)}{(z^2-c^2)^{3/2}} ,\qquad V'(z) = z - \sum_k g_k z^{2k-1},
 \eeq
 where the contour $C$ encloses the cut $[-c,c]$ on the real axis and where we assume that only a finite, but in principle 
 arbitrarily large, number of the $w_k$ can be different from zero .
 Finally, the cut  $[-c,c]$ is determined as a function of $g$ by the following equation for $c^2(g)$
 \beq\label{j8}
 \oint_C \frac{dz}{2\pi i} \; \frac{z V'(z)}{(z^2-c^2(g))^{1/2}} = 2.
 \eeq
 We present these explicit formulas because they tell us  how to take the continuum limit of the lattice theory. We are interested
 in a limit where the number $N$ of polygons goes to infinity. To each polygon we associate
 an indeterminate $g$, and  $Z(g,z_i)$ has a convergent power 
 expansion in $g$ for small $g$. Large $N$ will dominate when one reaches the radius 
 of convergence of $Z(g,z_i)$. This occurs either when $M_1(c^2(g),g) = 0$ or when $c^2(g)$ ceases to be an 
 analytic function of $g$. This happens to be at the same point $g_0$. This $g_0(w_j)$ will be a function 
 of the relative weights $w_k$, which in this discussion of convergence we consider fixed . 
 Similarly, we might be interested in the situation
 where the lattice lengths $l_i$ go to infinity. 
 This happens by the same reasoning when $z_i = c(g)$, where \rf{j6}
 is non-analytic in $z_i$. Denoting $z_0 = c(g_0)$, 
 $g_0(w_j)$ and $z_0(w_j)$  are critical points of our statistical system of graphs. By approaching these critical points according to
 \beq\label{j9}
 g  = g_0(w_j)  \, \e^{-\Lam a^2} = \e^{-\mu}, 
 \qquad \frac{1}{z} = \frac{1}{z_0(w_j)} \, \e^{-Z_i a}= \e^{-\lam_i}
 \eeq
 we can take the continuum limit of \rf{j6} by scaling $a \to 0$, and make contact with \rf{j4}:
 \beq\label{j10}
 Z(g,z_i) \propto a^{5-\frac{7n}{2}} \Big(\!- \!\frac{d}{d \Lam}\Big)^{n-3} 
 \Big[ \frac{1}{\sqrt{\Lam}} \prod_{k=1}^n \frac{1}{(Z_i\!+\! \sqrt{\Lam})^{3/2}}\Big] 
 \propto   a^{5-\frac{7n}{2} }Z(\Lam,Z_i),
 \eeq
valid for $n \geq 3$. In particular we find from \rf{j10} and \rf{j4} by inverse Laplace transformation
 \beq\label{j11}
Z(V,L_i) \equiv  \cN(V,L_i) \propto V^{n -\frac{7}{2}} \sqrt{L_1 \cdots L_n} \; \e^{- {L_1 + \cdots + L_n)^2}/{4V}}.
 \eeq
 This formula is also valid for $n=0,1,2$. 
 From \rf{j9}, \rf{j5a} and \rf{j5b} the relation between $\cN(V,L_i)$ and $\cN(N,l_i)$ is 
 \beq\label{j12}
 \cN (N,l_i) \propto \e^{\mu_0 N + \lam_i^{(0)} l_i}    \cN(V,L_i),\qquad \e^{\mu_0} = g_0,\quad \e^{ \lam_i^{(0)}} = \frac{1}{z_0},
 \eeq
 which shows that the number of generalized triangulations (bipartite graphs) with spherical topology and $n$ boundaries {\it grows exponentially} with $N$, the number of polygons in the graphs. The number of graphs also grows exponentially with $l_i$,
 the number of boundary links.
 Eq.\ \rf{j9} can be seen as  {\it additive renormalizations} of the cosmological and boundary cosmological constants $\Lam_0$ and 
 $Z^{(0)}_i$:
 \beq\label{j13}
\mu =  \Lam_0a^2 = \mu_0 + \Lam a^2, \qquad \lam_i = Z^{(0)}_ia = \lam_i^{(0)} +  Z_i a.
 \eeq
  
 The Wilsonian aspect of the above formulas is the following: we have an infinite-dimensional
 coupling constant space corresponding to $g_k \geq 0$. The critical surface is  defined by $g_k = g_0(w_j) w_k$, where for 
 given $w_k$, the $g_0(w_j)$ is the critical point discussed above. Thus the critical surface has co-dimension 1 and approaching it 
 for fixed $w_k$ like in \rf{j9} leads to the same continuum theory. In this sense it is a beautiful example of Wilsonian universality,
 but one can ask: where is the divergent correlation length in the lattice theory, leading to this universality?
 This is especially interesting since this is a theory of quantum gravity, and as discussed above, we are integrating 
 over to geometries which define length. We will discus this in the next subsection. 
 
 Let us end this subsection with a remark about the critical surface. 
 We have restricted $w_k$ to be larger than or equal to zero
 and with only a finite number of the $w_k$ different from zero. If one relaxes these constraints, in particular the constraint that
 the $w_k$ have to be positive, one can obtain different critical behaviors \cite{kazakov,abm} 
 (which can be given the interpretation of matter 
 systems coupled to quantum gravity). One obtains then a picture where  a fine-tuning of the bare coupling constants 
 to reach the critical surface might lead to different regions corresponding to different continuum theories. 
 We will not pursue this possibility any further in this Chapter, but only mention that the corresponding continuum quantum field theories 
 are the  so-called quantum Liouville theories with different central charge. These Liouville theories 
 arise when quantizing two-dimensional Euclidean gravity coupled to conformal matter. Integrating out the matter fields, while 
 using the conformal gauge for the metric leads to an effective quantum field theory for the conformal factor of the metric, the Liouville
 quantum field theory, which depends on the central charge of the conformal 
 matter field integrated out \cite{kpz,davidconformal,dk1,dk2}. 
 The relation between the central charge $c_L$ of the Liouville theory 
 (which is also a conformal theory, although a somewhat special one) and the central charge $c$ 
 of the matter field is $c_L = 26 - c$. The pure two-dimensional
 quantum gravity theory we have mainly discussed above corresponds in this notation to a conformal theory with central 
 charge $c=0$ and thus a Liouville theory with $c_L=26$. The last 20 years have seen major progress in 
 understanding and formulating the mathematics behind Liouville quantum gravity, and Chapter 7 in this Section
 of the Handbook, ``Lessons from Mathematics of Two-dimensional Quantum Gravity",  
 will describe this in detail.

 \subsubsection*{A divergent correlation length in 2d EDT}
 
 In an ordinary quantum field theory in flat spacetime a correlator is defined by 
 \beq\label{j14}
 \la \phi(x) \phi(y) \ra = \frac{\int \cD \phi \;  \e^{-S[\phi]}\;\phi(x) \phi(y) }{\int \cD \phi \, \e^{-S[\phi]}}
 \eeq
 By translational and rotational invariance 
 (which we will asssume) $\la \phi(x) \phi(y) \ra $ is  only  a function 
of $|x-y|$, where $x$ and $y$ are spacetime points. We can take advantage of this by averaging over all points
 $x$ and $y$ separated by a distance $|x-y| = R$ and define 
 \beq\label{j15}
   \la \phi \phi \ra_R  = \frac{\int \cD \phi \; \e^{-S[\phi]} \; \int dx \int dy \;\del (|x-y| - R) \; \phi(x) \phi(y) }{ \int \cD \phi \; \e^{-S[\phi]}},
 \eeq
 where formally this average contains a factor  $V$, the volume of spacetime, due to translational invariance. 
 We can embed this definition of a correlation function in a quantum gravity theory
 \beq\label{j16}
  \la \phi \phi \ra_R  =
  \frac{\int\cD [g] \cD \phi \; \e^{-S[g,\phi]} \; \int dx \int dy \sqrt{g(x) g(y)} \;\del (D_g(x,y) - R)  \,
  \phi(x) \phi(y) }{ \int \cD[g]\cD \phi \; \e^{-S[g,\phi]}},
 \eeq 
 where $S[g,\phi]$ denotes the combined action of gravity and the field theory, and where $D_g(x,y)$ is the geodesic distance 
 between spacetime points labelled $x$ and $y$. The correlation function \rf{j16} is diffeomorphism-invariant, but non-local.
 Since the gravity action contains a cosmological constant $\Lam$, the average volume of spacetime will be finite and 
 of order $1/\Lam$. Contrary to \rf{j15} there is no infinite formal factor in the definition \rf{j16}. We call $R$ the 
 {\it quantum geodesic distance}. Note that it will influence $\la \phi \phi \ra_R$ in a 
 potential much more  radical way than the 
 $R$ in \rf{j15}; when $R$ is large compared to some appropriate power of $1/\Lam$, it will define the shape of 
 the whole universe in which we measure the correlation\footnote{\label{elongated} 
 For such a 
 large $R$, the universe will be quite ``elongated'', because by definition at least two 
 points have to be separated a geodesic distance $R$.}. Thus, in some ways $R$ is more like a new coupling constant 
 in the theory, in the sense that the average shape of the universe depending it. 
 
\begin{figure}[t]
\centerline{\scalebox{0.17}{\includegraphics[angle = 0]{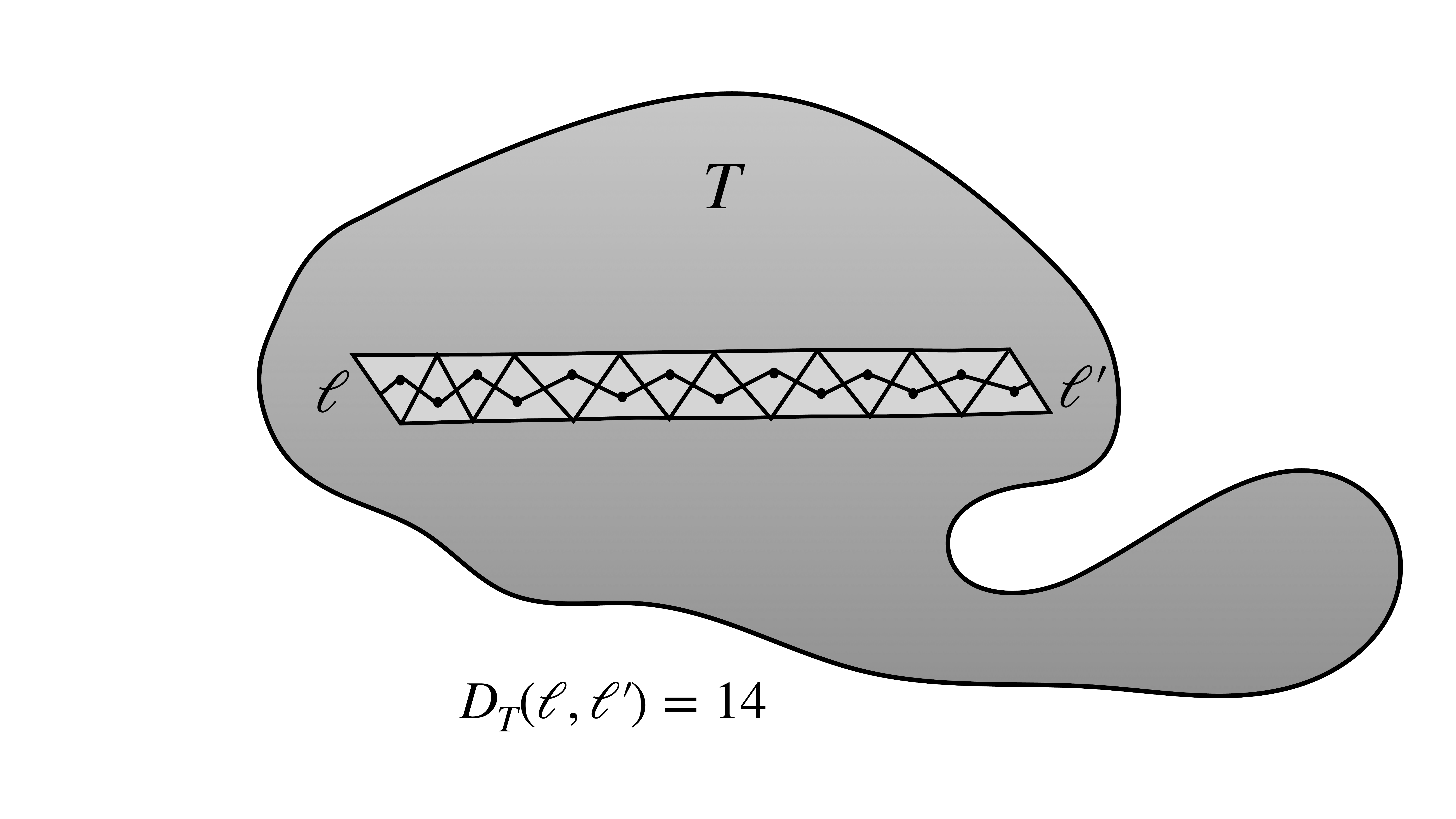}}}
\caption{{\small A triangulation $T$ with two links $\ell$ and $\ell'$ separated by a graph distance $D_T(\ell,\ell') = 14$.}}
\label{figj1}
\end{figure}

 In the case of pure gravity, one has no external field $\phi(x)$, but one could consider curvature--curvature correlators\footnote{Chapter 2 in this Section of the Handbook,
 ``Observables and Curvature in CDT'', 
 describes how to introduce curvature in lattice gravity theories.}, or simply replace 
 $\phi(x)$ by $1(x)$, which takes the value 1 for all $x$. 
 This last choice is  of interest since the correlator has a clear geometric interpretation
 and  it can be explicitly calculated in the two-dimensional lattice gravity theory. We 
 define the (unnormalized) two-point 
 function corresponding to \rf{j16} with $\phi(x) = 1(x)$ as 
 \beq\label{j17}
 G_\Lam(R) = \int\cD [g]  \; \e^{-\Lam V_g} \; \int dx \int dy \sqrt{g(x) g(y)} \;\del (D_g(x,y) - R) 
 \eeq 
 This is a formal continuum definition and requires a UV cut-off to define the path integral in \rf{j17}. Again we use EDT and 
 in addition  now have to define the geodesic distance between the spacetime points $x$ and $y$ in formula \rf{j17}
 in the context of our triangulations. Let us for simplicity  consider
  a triangulation constructed from equilateral triangles.  As already mentioned, this 
 triangulation can be viewed  as a piecewise linear surface where the geometry is uniquely defined 
 by assuming the triangles are flat in the interior. 
 From such a piecewise linear triangulation, where one knows the length of each link,
 one can calculate $D_g(x,y)$. However, an approximate definition, convenient from a calculational  
 point of view, is to define the graph distance between two links\footnote{One could also have chosen to define the 
 graph distance between two vertices as the shortest link distance between the two vertices. As usual, from a Wilsonian 
 point of view one should be led to the same continuum limit if it exists.} as the shortest distance,
 passing through centers of neighboring triangles, see Fig.\ \ref{figj1}. 
 
 For generic, very large triangulations, and for 
 links with correspondingly large separation, we expect such a distance to be proportional
 to  the ``real'' geodetic distance. Denote this graph distance between link $\ell$ and link $\ell'$ in a triangulation $T$ 
 of the sphere by $D_T(\ell,\ell')$.
  The  lattice equivalent of \rf{j17} can be written as  
 \beq\label{j18}
 G_\mu(r) = \sum_{T} \e^{-\mu N(T)} \sum_{\ell,\ell'} \del_{D_T (\ell,\ell'),r}.
 \eeq
 Quite remarkably, one can combinatorially calculate the sum 
 of these triangulations \cite{aw}. Close to the critical point $\mu_0$ 
 defined in \rf{j12} one obtains
 \bea\label{j19}
 G_\mu (r) &\propto& (\mu-\mu_0)^{3/4} \; \frac{\cos \sqrt[4]{\mu-\mu_0} \, r}{\sin^3 \sqrt[4]{\mu -\mu_0} \, r} \\
  G_\mu(r) &\propto& (\mu-\mu_0)^{3/4} \; \e^{ - 2\sqrt[4]{\mu -\mu_0} r}\quad {\rm for}\quad  r \gg  \sqrt[4]{\mu -\mu_0},\label{j19a}
 \eea
 i.e. an exponential fall-off with a correlation length $\xi(\mu) = 1/ \sqrt[4]{\mu -\mu_0} $.
 Using \rf{j13}, we can directly read off the continuum limit of \rf{j19},  {\it provided the geodesic distance $R$ scales 
 anomalously}: 
  \beq\label{j20}
   a^{-3/2} G_\mu (r)  \propto   G_\Lam(R) =  \Lam^{3/4}  \frac{\cos \sqrt[4]{\Lam} \, R}{\sin^3 \sqrt[4]{\Lam} \, R},
   \qquad R = a^{1/2} r.
   \eeq
  Note that
    the anomalous dimension of $R$ shows that the two-dimensional EDT quantum spacetime is fractal, with Hausdorff 
   dimension $d_h =4$ at all scales, as first realized in the seminal work \cite{kawai-wata}. Furthermore,
   \bea\label{j21}
   \chi(\mu) &=& \sum_{r=1}^\infty G_\mu(r)  = {\rm const.} - \frac{1}{6} \sqrt{\mu-\mu_0} + O(\mu-\mu_0)  + \cdots \\
   &\equiv& {\rm analytic} + 
   \frac{1}{(\mu - \mu_0)^\gamma} + \cdots.
   \eea
  where  $\chi(\mu)$  denotes the susceptibility. The term  $(\mu-\mu_0)^{-\gamma}$ is the leading non-analytic term
   in the expansion of $\chi(\mu)$ around $\mu_0$, and $\gamma$ is called the susceptibiliy exponent. These notations
   are inspired by the analogous notations used for spin-spin correlation functions in the theory of critical phenomena. 
   From the definition \rf{j18}
   it follows that $\chi(\mu) \propto d^2 Z(\mu)/d\mu ^2$, where $Z(\mu) $ is   given by \rf{j5a} with $n=0$ (no boundaries).
    We thus obtain
   \beq\label{j22}
   Z(\mu) \equiv {\rm analytic} + (\mu - \mu_0)^{2-\gamma} + \cdot , \qquad \gamma = - \oh,
   \eeq
   a result which is consistent with \rf{j11} and \rf{j12}. We have identified a divergent correlation length of two-dimensional  EDT, and 
   it is directly related to the fractal structure of the corresponding spacetime. The existence of 
   this divergent correlation length explains why the Wilsonian picture works so well in this model. 
    A final remark concerns the quantum geodesic distance $R$ which appears in the definition \rf{j17}. Eq.\ \rf{j20} shows 
    how the choice of $R$ will affect the general shape of the universe (as already mentioned 
    in footnote \ref{elongated}): for $R \gg 1/\Lam^{1/4}$ it is a long 
    tube of length $R$ and cross-section proportional to $\Lambda^{-3/4}$.

 \subsubsection*{The generalized two-dimensional CDT theory} 
 
\begin{figure}[t]
\centerline{\scalebox{0.17}{\includegraphics{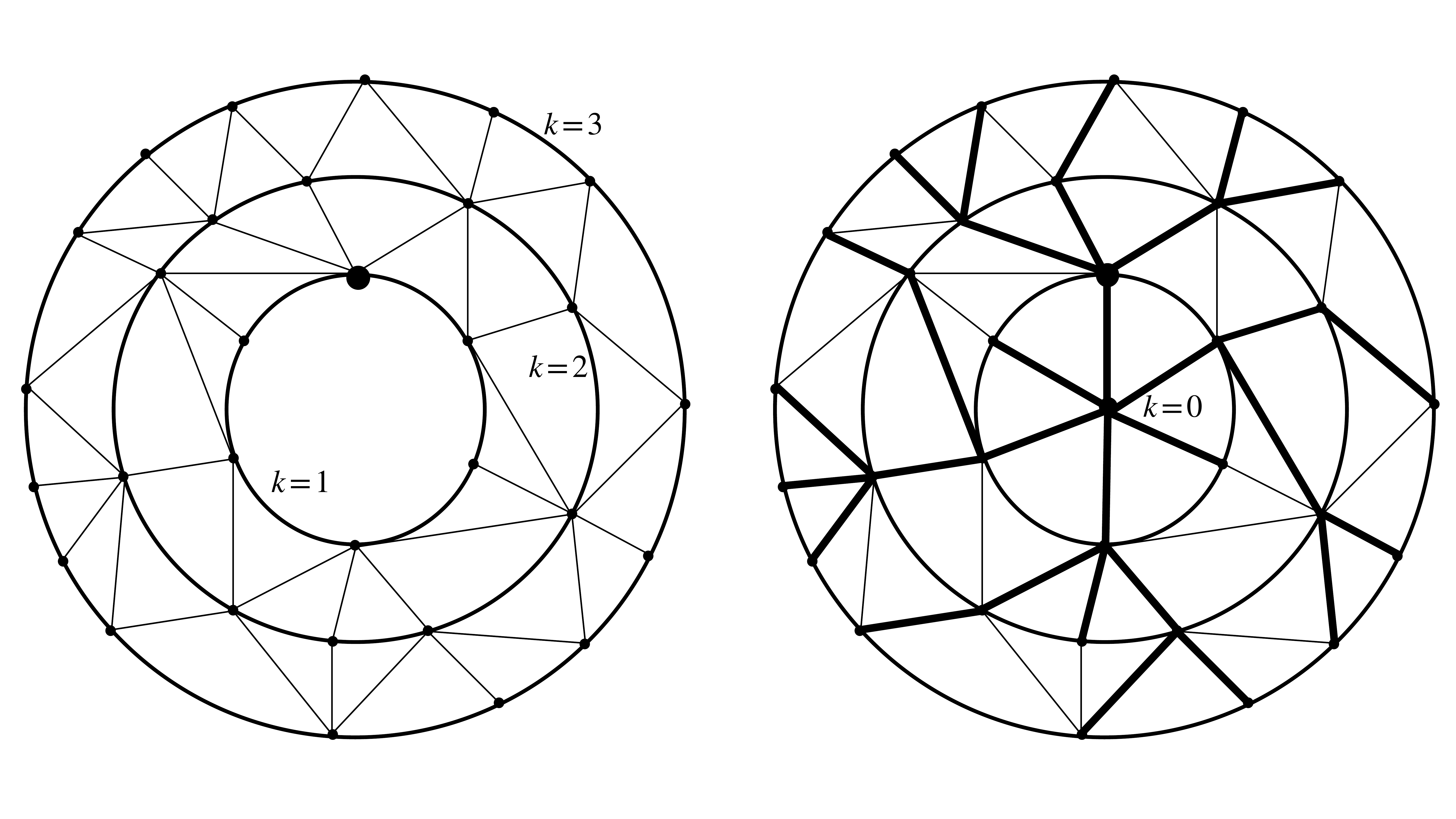}}}
\caption{{\small  Left figure: A CDT triangulation  (represented as an annulus). Constant time slices corresponding to 
$k=1,2,3$ are circles.  
A vertex (or the spatial link to the right of it) on the entrance loop $k\equ 1$ is marked. Right figure: 
the corresponding branched polymer (thick black links). 
An artificial vertex at $k \equ  0$ connected to each vertex at the $k \equ 1$ loop 
ensures a bijection between the CDT triangulations with boundaries at times $k=1$ and $k=3$ 
and so-called rooted  branched polymers of {\it height $3$} (the root connects the vertex at $k \equ 0$ to the marked 
vertex at $k \equ 1$).}}
\label{figj2}
\end{figure}

 As discussed above, the scaling limit for 2d EDT is essentially independent  of the choice of weight $w_n$ of the polygons, as long as 
 the weights are non-negative. In a Wilsonian context, a change of universality class is most likely related to a change of 
 some global symmetry. The EDT formalism respects in a formal way the symmetry between space and (Euclidean) time, to a 
 degree that it is unclear how one would actually rotate expressions like the two-point function $G_\Lam(R)$ to spacetimes with 
 a Lorentzian signature. Two-dimensional Causal Dynamical Triangulations (CDT) is a regularization 
 which takes the difference between space and time 
 serious from the outset, and insists on summing over spacetimes which have a well-defined time foliation. It is simplest to
 implement this in a discretized path integral if one assumes that space has the topology of a circle. Two neighboring
 spatial slices at discretized integer  times $k$ and ${k+1}$ then consist of $l_k$ and $l_{k+1}$ spatial links, and the two 
 slices are connected by triangles with one spatial link and two time-like links, in such a way that the corresponding 
 two-dimensional triangulation with the spatial slices at $k$ and ${k+1}$ has the topology of a cylinder, as illustrated in 
 Fig.\ \ref{figj2}. Clearly, one can in this way iteratively construct a two-dimensional triangulation 
 with spatial slices at $k$, $k=1,\ldots,s$, consisting of $l_k$ links. This yields
   a cylinder with an ``entrance'' spatial loop 
 consisting of  $l_1$ and an ``exit'' spatial loop consisting of $l_s$ links, as also shown in Fig.\ \ref{figj2}.
 Like in the EDT case, only the cosmological term will be important if we  sum over
 piecewise linear manifolds with a fixed topology  in the path integral. We can write, for a Lorentzian triangulation $T_{\rm lor}$ of the kind discussed
 \beq\label{j23}
 S_{T_{\rm lor}}(\Lam,\al)  = -\Lam N(T_{\rm lor}) \frac{\sqrt{4\al +1}}{4}a^2,  \qquad a^2_t = -\al a^2, \quad \al >0.
 \eeq 
 In \rf{j23} we use the explicit area of a triangle with one spatial link of length $a_s^2 = a^2$ and two time-like links with 
 $a_t^2 = - \al a^2$ (see \cite{physrep,ajl} for details). If $\al > 1/4$ we can perform an analytic continuation in the lower complex $\al$-plane to negative $-\al$ such that
 \beq\label{j24}
 S_{T_{\rm lor}} (\Lam,\al) \to S_{T_{\rm lor})}(\Lam,-\al-i\ep) = i S_{T_{\rm eucl}} (\Lam,\tilde{\al}) ,
 \eeq
 where the Eucledian triangulation is denoted $T_{\rm eucl}$, and where
 \beq\label{j24a}  
 S_{T_{\rm eucl}} (\Lam,\tilde{\al}) =  \Lam \;N(T_{\rm eucl}) \;\frac{\sqrt{4 \tilde{\al} -1}}{4} a^2\qquad \tilde{\al} = \al > \frac{1}{4}.
  \eeq  
  The inequality $\tilde{\al} > 1/4$ has the simple geometric interpretation that the sum of lengths of the two ``time-like" triangle sides
 (i.e.\ $2 \sqrt{\tilde{\al} }\;a$)  has to be larger than the length $a$ of the space-like side of a triangle in the ``flat''  Euclidean triangles
 used in the rotated triangulation.  
  
  Eq.\ \rf{j24} is the ``usual'' formal relation between the Lorentzian   and Euclidean actions, such that 
  \beq\label{j24b}
  \e^{iS_{T_{\rm lor}}(\Lam,\al)} = 
  \e^{-S_{T_{\rm eucl}}(\Lam,\tilde{\al})},\qquad \tilde{\al} = \al > \frac{1}{4}.
  \eeq
  For each $T_{\rm lor}$ we perform the rotation to a corresponding $T_{\rm eucl}$ with the actions related by \rf{j24}.
  The important point is that the  class of triangulations $\{ T_{\rm eucl}\}$ obtained in this way is quite different from the 
  class used in EDT.  From now we will set $\al=1$ since it only contributes a constant of proportionality
  to the action, where we have anyway already absorbed a factor of proportionality in $\Lam$. 
  We thus write, as in EDT\footnote{Again we 
  assume, as in the EDT case, that a boundary link is marked on one of the boundary loops, to avoid symmetry factors occuring
  in the sum over triangulations.}
  \beq\label{j25}
  S_T(\Lam) = \Lam N(T) a^2  = \mu N(T), \qquad \mu = \Lam_0 a^2,
  \eeq
  \bea\label{j26}
   Z(\mu,\lam_1,\lam_s) &=& \sum_T \e^{ -\mu N(T) -  \lam_1 l_1 - \lam_s l_s}\\
   &=& \sum_{N,l_1,l_s} \e^{ -\mu N(T) -  \lam_1 l_1(T) - \lam_s l_s(T)}
   \cN(N,l_0,l_s), \label{j26a}
   \eea
   where the summation is over the triangulations described above, which have the topology of a cylinder, with 
   $s$ spatial slices, where slice 1 consists of $l_1$ spatial links and slice $s$ of $l_s$ spatial links. 
 
  The two-dimensional CDT  model (and related models) can be solved analytically \cite{al,charlotte,ai}, 
 and rather surprisingly the critical exponents of the model agree 
 with corresponding  critical exponents of tree graphs or so-called
 branched polymers. Later it was understood that this  is not a coincidence \cite{djw}, but that there exists a bijective map of 
 the CDT surface graphs onto so-called rooted branched graphs of height $s+1$, as illustrated in Fig.\ \ref{figj2}.
 This insight  highlights  the importance of tree-like structures in graphs relevant to quantum gravity. Chapter 5 in this 
 Section of the Handbook, ``From Trees to Gravity",  is  dedicated the study of such tree-like graphs.
 
 If we define a slightly modified CDT graph by connecting all vertices at time-slice 
 $k=1$ to a single vertex at a new time-slice at $k=0$
 and all vertices at time-slice $k=s$ to a single vertex at a new time-slice at $s+1$ then the graph distance (which we here 
 define to be the shortest link distance) between any vertices
 will be less or equal  to $s+1$. If we start at the vertex at time 0, then the only vertex  where the graph distance to the starting 
 vertex is a  local maximum is the vertex at $s+1$ (and the local maximum is in this case also a global maximum). From a graph point of view this is a rather special situation and 
 one can generalize it to include graphs where a finite number of vertices have a local maximum distance to a starting vertex,
 even in the limit where the number of vertices  goes to infinity. This is the setup of generalized CDT: starting from a vertex or 
 a spatial entrance loop, one moves forward in ``proper time'', which is defined as the graph distance from the vertex or the 
 entrance loop (and in the continuum by the geodesic distance from the entrance loop). On the way to the exit spatial loop (or loops),
 space can branch into several disconnected spatial universes. 
 The ones that do not end in exit loops vanish into the 
 ``vacuum'' . The distances of these vacuum points to the entrance loop are then local maxima, and the spatial loops 
 that in this way disappear into the vacuum are called  baby universes. 
 For graphs consisting of a finite number of vertices, there is no real difference
 between the graphs used in EDT and the ones used in generalized CDT, 
 but the crucial difference comes from 
 requiring that when the number of vertices goes to infinity
 the number of baby universes stays finite. This will then also be true in the  continuum limit and  is  in contrast to the 
 EDT situation, where the fractal structure with Hausdorff dimension $d_h=4$ implies that infinitely many baby universes (but most of them with infinitesimal 
 volume) are created in the continuum limit. 
 In the case of generalized CDT one finds $d_h=2$. Again the discretized model can be solved analytically 
 and one can take the continuum limit\footnote{Rather amazingly, it is possible to solve the model
 directly in the continuum simply be using that the number of baby universes is finite \cite{alwz1,alwz2,alwz3}} in much the same way as was done in the EDT case \cite{ab}. 
 In particular, one can find the continuum
 version of the generalized CDT two-point function
 \beq\label{j27}
 G_\Lam(\tau) = \frac{\Sigma^3}{\Theta}\; 
 \frac{ \Sigma \sin \Sigma \tau + \Theta \cos \Sigma \tau}{\Big(\Sigma \cos \Sigma \tau + \Theta \sin \Sigma \tau\Big)^3}
 \eeq 
 where 
 \beq\label{j28}
 \Sigma = \sqrt{\Lam} \;H\Big( \frac{G_b}{\Lam^{3/2}}\Big),\quad H(0)=1; \qquad  
 \Theta = \sqrt{\Lam}\; F \Big( \frac{G_b}{\Lam^{3/2}}\Big), \quad F(0) = 1.
 \eeq 
 The functions $H(x)$ and $F(x)$ have a power expansion in $x$, with a radius of convergence ${2}/{3^{3/2}}$. A new 
 coupling constant denoted $G_b$ has appeared in \rf{j28}. It  is the coupling constant for a spatial universe to split into two spatial 
 universes. When $G_b/ \Lam^{3/2}\to 0$ we get back to ordinary CDT. When 
 $G_b/\Lam^{3/2} \to 2/3^{3/2}$, \rf{j27} and \rf{j28} cease to be valid and one can show that the number of baby universes goes to 
 infinity, indicating that one has a phase transition to ordinary EDT gravity. We have in \rf{j27} denoted the geodesic distance
 entering in the two-point function by $\tau$ (proper time) rather than the $R$ used in \rf{j20}, to emphasize the origin 
 as a proper time in Lorentzian CDT. Eq.\ \rf{j27} looks 
 superficially like  a generalization of eq.\ \rf{j20}. However, the important point is 
 that we have $\sqrt{\Lam}\, \tau$ as an argument, while in \rf{j20} $\sqrt[4]{\Lam}\, R$ appears as an argument, capturing 
 the difference in Hausdorff dimension for the two ensembles of geometries. We still have a perfect Wilsonian picture 
 for generalized CDT as embedded in EDT \cite{cdtmatrix1,cdtmatrix2}. 
 In  EDT one can introduce an additional dimensionless coupling $g_b$,
 which controls the creation  of baby universes, such that for  small values of $g_b$  the creation of baby universes is suppressed.
  One can show that for any finite value of $g_b$ the critical behavior of the system is still that of two-dimensional 
   Euclidean gravity. However,  if 
 $g_b$ is scaled to zero at the same time as one approaches the critical surface in the following way (which is a 
 generalization of \rf{j13})
 \beq\label{j29}
 g_b = G_b a^3, \quad \mu = \mu_0 + \Lam a^2,  \quad \lam_i = \lam^{(0)}_i + Z_i a,
 \eeq
 one obtains generalized CDT with continuum coupling constants $\Lam$, $Z_i$ and $G_b$.
 From a Wilsonian point of view we have an infinite-dimensional critical surface and on this critical surface a subspace 
 where $g_b =0$. On this  critical subspace 
 there is an asymmetry between  ``space'' and ``time'', the same asymmetry that 
 was put in by hand in the original simple CDT model and taht was not present in the EDT model. 
 Approaching the subspace as in \rf{j29} one will obtain the continuum limit corresponding to generalized 
 CDT, while approaching the critical surface at a point where $g_b > 0$, in the way described by \rf{j13} leads to the continuum 
 limit of EDT.
 
 Ho\v{r}ava-Lifshitz gravity is a  continuum theory where one demands that spacetime has a time foliation and that the theory is invariant
 under spatial diffeomorphisms and time redefinitions \cite{horava}. 
 Chapter 4 in this Section of the Handbook discusses (among other topics) 
 the relation between 
 two-dimensional CDT and two-dimensional Ho\v{r}ava-Lifshitz gravity.
 (see also \cite{agsw}). In the case of higher dimensional CDT no such 
 relation is known to exist.

 \section{Four-dimensional EDT}
 
 While two-dimensional EDT and CDT have a clear Wilsonian interpretation where  continuum limits can be defined 
 and continuum correlators can be calculated analytically, the situation is more complicated when one wants to 
 generalize the EDT and CDT formalism to higher-dimensional gravity.  
 First, higher-dimensional gravity is non-renormalizable,
 and the curvature term that dropped out in two-dimensional quantum gravity is now expected to play a key role. In the Wilsonian 
 context of asymptotic safety\footnote{The concept of asymptotic safety and the way it is implemented in the case of gravity is the topic of the Chapter called ``Asymptotically Safe Quantum Gravity" in the Handbook.}, one needs a non-trivial UV fixed point of the lattice theory if conventional renormalization 
 group logic applies and if one wants the lattice theory to define a quantum continuum theory at all scales. Secondly, 
 contrary to the situation in two dimensions, there is presently no way we can solve the lattice theory analytically. We have 
 to rely on Monte Carlo simulations of the path integral. This implies that we have to use an action with Euclidean signature,
 since the Monte Carlo simulations need the exponential of the action to have a probability interpretation.
 Unfortunately, the Euclidean four-dimensional  continuum Einstein-Hilbert action
 is unbounded from below and this will be true also for the lattice action 
 when the lattice volume becomes infinite. Only the measure term in the path integral may save us, if we want to restrict ourselves
 to  the  Einstein-Hilbert term as the classical action appearing in the path integral. Alternatively, one could include higher curvature 
 terms in the classical action. Finally, we have to be able to find second or higher order phase transitions for the lattice theory, as 
 discussed above.
 
 Let us define the lattice theory. We follow the 
 two-dimensional theory and consider four-dimensional piecewise linear geometries, constructed by gluing together 
 building blocks consisting of four-simplices, where all link lengths are equal to $a$, 
 which then serves as our UV cut-off. The only 
 restriction on the gluing is that the gluing locally is such that one  has a (piecewise linear) manifold and that 
 the topology of this piecewise linear manifold is fixed \footnote{One can study more general models where one 
 relaxes the constraint that the gluing should result in a piecewise linear geometry or that the topology of the mainfolds
 should be fixed. It is possible to formulate such a generalized gluing procedure in different ways and starting with the 
 articles \cite{adj,sasakura}, these models are denoted tensor models. If we discuss the gluing of $d$-dimensional simplices, 
 the number of tensor indices are equal to the number of $d-1$ dimensional subsimplices in the $d$-dimensional simplices  
 which constitute the building blocks. For $d=2$ we have tensors of rank 2, i.e.\ matrices, and two-dimensional 
 gravity has indeed been studied using matrix models, starting with the work of David \cite{david}.}. 
 Most studied is the simplest topology, the four-sphere $S^4$,
 and we will limit ourselves to discussing this case. We consider the four-simplices as flat in the interior. View as a piecewise linear continuum manifold all geodesic 
 distances are well-defined, and thus  the geometry of such a triangulation 
 is  also fixed without specifying a coordinate system. Summing over combinatorially inequivalent triangulations then results in a summation over a certain class
 of piecewise linear geometries, and the hope is that in the limit where
  the lattice spacing $a \to 0$, this summation will in some sense
 be a good representation of the integration over continuous geometries (of which the piecewise linear geometries constitute
 a subset, that is hopefully dense with respect  to a (still) unknown measure).  
 
 An obvious question  is how to represent the curvature term present in the Einstein-Hilbert action. Regge showed
 how to define curvature locally on a $d$-dimensional piecewise linear manifold \cite{regge} constructed from $d$-simplices $\sg^d_j$, 
 by locating it on $(d-2)$--dimensional subsimplices $\sg^{d-2}_i$. 
 The $d$-simplices $\sg^d$ sharing a $(d-2)$--dimensional subsimplex $\sg^{d-2}$, have 
 dihedral angles\footnote{For a given $d$-simplex, any of its $(d-2)$-subsimplices will be the intersection of precisely two 
 of its $(d-1)$--subsimples, and the angle between these two $(d-1)$--subsimplices is called the dihedral angle (which {\it is} an angle 
 for any $d \geq 2$).} $\th(\sg^{d-2},\sg^d)$, related to this subsimplex.  
 If the spacetime was flat these dihedral angles would add to $2\pi$. The so-called 
 deficit angle $\eps_{\sg^{d-2}}$, associated with the subsimplex $\sg^{d-2}$  is defined by 
\beq\label{j30}
 \eps_{\sg^{d-2}} = 2\pi\;\; - \hspace{-5mm}\sum_{\{\sg^d | \sg^{d-2} \in \sg^d\}} \th( \sg^{d-2},\sg^d).
 \eeq
 The deficit angle is the  angle by which a vector will be rotated when parallel-transported locally around the $(d-2)$--simplex in 
 the piecewise linear geometry in the subspace perpendicular to the $d-2$-dimensional
 simplex. Regge  showed that the curvature action and the volume term 
 associated to the piecewise linear manifold $M$ can be written as 
 \beq\label{j31}
 \int_{M} \d^d \xi\; \sqrt{|g(\xi)|}\; R(\xi) = 2 \sum_{\sg^{d-2}} \eps_{\sg^{d-2}} V_{\sg^{d-2}}, \qquad  
 \int_{M} \d^d \xi\; \sqrt{|g(\xi)|} = \sum_{\sg^d} V_{\sg^d},
 \eeq 
 where $V_{\sg^{d-2}}$ denotes the volume of the subsimplex $\sg^{d-2}$ and  $V_{\sg^{d}}$ the volume of the simplex $\sg^d$
 (in the case there $d=2$ we define  $V_{\sg^{d-2}}=1$).
 
 For the piecewise linear geometries used in EDT this expression simplifies enormously since all dihedral angles are identical, 
 all $(d-2)$--volumes are the same and all $d$-volumes are also equal. In the four-dimensional case, which has our
 main interest, we have 
 \beq\label{j31a}
  \th( \sg^2,\sg^4) = \arccos \oq,\qquad V_{\sg^2} =\frac{\sqrt{3}}{2} a^2, \qquad  V_{\sg^4} =\frac{\sqrt{5}}{96} a^4.
  \eeq
 The Einstein-Hilbert action for a given triangulation $T$ of a closed four-dimensional manifold in EDT can then be written as 
 \bea\label{j32}
 S_M (G, \Lam) &=& \frac{1}{16\pi G} \int d^4\xi \sqrt{g(\xi)} \;\big (-R(\xi) + 2\Lam\big) \to \nonumber \\
  S_T(\kp_2,\kp_4) &=& - \kp_2 N_2(T) + \kp_4 N_4(T),  
 \eea
 \beq\label{j32b}
  \kp_2 = \frac{1}{8G}\, \frac{\sqrt{3}a^2}{2},\qquad 
 \kp_4 =\frac{2\Lam}{16\pi G} \,\frac{\sqrt{5} a^4}{96}  + 20 \arccos \Big(\oq\Big)  \frac{1}{16\pi G} \frac{\sqrt{3} a^2}{2}, 
 \eeq
 where $N_2(T)$ denotes the number of two-simplices and $N_4(T)$ the number of four-simplices in the triangulation $T$.
 From the so-called Dehn-Sommerville relations for a closed four-dimensional 
 triangulation $T$ one has that $N_2(T) = 2N_0(T) + 2N_4(T) -2\chi$, where $N_0(T)$ denotes the number of vertices in the 
 triangulation and $\chi$ is the Euler characteristic of the triangulation. The Euler characteristic only depends on the 
 topology of the triangulation.  Thus \rf{j32} can be written as 
 \beq\label{j32a}
 S_T(k_0,k_4) = -k_0 N_0(T) + k_4 N_4(T) +  k_0 \chi,\qquad k_0 = 2\kp_2,\quad k_4 = \kp_4 - 2\kp_2
 \eeq 
 where the  $\chi$-term is usually ignored since we consider triangulations with a fixed topology. 
 
 The EDT partition function of four-dimensional quantum gravity is now obtained by summing over triangulations with the 
 action given by \rf{j32} (or \rf{j32a}):
 \beq\label{j33}
 Z(\kp_2,\kp_4) = \sum_T \frac{1}{C_T} \;\e^{\kp_2 N_2(T) - \kp_4 N_4(T)} = \sum_{N_4,N_2} \e^{ \kp_2 N_2 - \kp_4 N_4} \cN (N_2,N_4),
 \eeq
 where the summation is over all abstract triangulations\footnote{We use here the 
 notation ``abstract triangulation'' to emphasize that although we have viewed the 
 triangulations as piecewise linear manifolds and have introduced a link length $a$ as 
 a UV cut-off, in the summation \rf{j33} only the labelling as (abstract) triangulations is 
 important. The other aspects will be important when we discuss a continuum limit.} $T$ of a given four-dimensional manifold, where $C_T$ is a symmetry factor 
 (the order of the automorphism group  of the triangulation $T$), and where 
 $\cN(N_2,N_4)$ denotes the number of such triangulations with a fixed number of two-simplices and four-simplices,
 $N_2$ and $N_4$, respectively. As was the case in two dimensions, 
 {\it the partition function is entirely combinatorial}: 
 $Z(\kp_2,\kp_4)$ is the generating function (with indeterminates 
 $e^{\kp_2}$ and $e^{-\kp_4}$) for 
 the number of four-dimensional triangulations with a given topology (here $S^4$) and a given number of four-simplices and 
 two-simplices. It is truly remarkable that four-dimensional quantum gravity in this way is purely ``entropic''. Unfortunately, it is not yet possible to perform this counting analytically.
  This leaves us presently 
 with Monte Carlo simulations if we want to study the partition function \rf{j33}.
 
 In view of the unboundedness of the Euclidean Einstein-Hilbert action, 
  the first obvious question one can ask is whether 
 $Z(\kp_2,\kp_4)$ is at all well defined for any values of $\kp_2$ and $\kp_4$. Let us perform the summation over $N_2$ in \rf{j33},
 \beq\label{j34}
 Z(\kp_2,\kp_4)  = \sum_{N_4} \e^{-\kp_4 N_4} \cN_{\kp_2} (N_4), \qquad \cN_{\kp_2} (N_4) = 
 \sum_{N_2} \e^{\kp_2 N_2} \cN(N_2,N_4).
 \eeq
If $\cN_{\kp_2} (N_4) $ is exponentially bounded as a function of $N_4$, i.e. if there exists a constant $\kp_4^c(\kp_2)$ such that 
\beq\label{j35}
\cN_{\kp_2} (N_4) \leq \e^{(\kp_4^c(\kp_2) + \eps)N_4}, \quad{\rm for~all}~ \eps > 0,~~N_4 > N_4(\eps),
\eeq 
 then there is a line $(\kp_2,\kp_4^c(k_2))$ in the $\kp_2,\kp_4$ coupling-constant plane, 
 such that $Z(\kp_2,\kp_4)$ is well defined and convergent for $\kp_4 > \kp_4^c(\kp_2)$.
 It is easy to prove that if \rf{j35} is valid for $\kp_2 = 0$, then it is valid for all $\kp_2$ (with a $\kp_4^c(\kp_2$) depending 
 on $\kp_2$). However, there is no proof that $\cN_0(N_4)$ {\it is } exponentially bounded. Computer simulations indicate that 
 it is the case \cite{aj,bt} and in the following we will assume so. 
 The physics of \rf{j34} is all hidden in $\cN_{\kp_2} (N_4)$, which according to our assumptions can be written as 
\beq\label{j36a}
\cN_{\kp_2} (N_4) = \e^{\kp_4^c(k_2) N_4} H_{\kp_2}(N_4),\qquad H_{\kp_2}(N_4)~{\rm subleading~in~} N_4,
\eeq
which implies that the non-trivial continuum physics 
is to be found in the {\it subleading} function $H_{\kp_2}(N_4)$.
One then obtains
\beq\label{j36b}
 Z(\kp_2,\kp_4)  = \sum_{N_4}  \e^{-(\kp_4 - \kp_4^c(k_2)) N_4} H_{\kp_2}(N_4).
 \eeq

 Given two triangulations $T(N_4)$ and $T(N_4')$ there exist local changes in the 
 triangulation $T_4$ (the so-called Pachner moves) that, when applied  a finite number of times, will bring us from 
 $T(N_4)$ to $T(N_4')$. These Pachner moves \cite{pachner} are used in the Monte Carlo simulations, 
 and allow us to have a Monte Carlo 
 algorithm that is ergodic  and  in principle creates the correct distribution Boltzmann distribution for \rf{j33}, corresponding to 
 the action \rf{j32}. Details will be provided in other Chapters of this Section 
 of the Handbook, see ``Spectral Observables and Gauge Field Couplings in Causal Dynamical Triangulations", Chapter 3, and ``Semiclassical and Continuum Limits of 
 Four-Dimensional CDT", Chapter 9.
 . 
 However, an interesting aspect in four dimensions is that  $N_4$ cannot 
 be kept fixed in these 
 Pachner moves, and it is even 
 impossible  in principle to calculate the highest $\tilde{N}_4 (T(N_4),T(N_4'))$ of an intermediate 
 triangulation $T(\tilde{N}_4)$ that one meets when moving 
 from the triangulation $T_4(N_4)$ to $T_4(N'_4)$ by successive application of the Pachner moves \cite{aj1}. It is unclear what this  implies 
 for the practical ergodicity of the used Monte Carlo simulations. It does not necessarily imply that $\tilde{N}_4$ is very large,
 but we cannot {\it in principle} provide a general 
 expression for $\tilde{N}_4 (T(N_4),T(N_4'))$. 
 
In the following we will assume that the Monte Carlo simulations work fine, despite the potential problems mentioned above. 
In the region of coupling-constant space $(\kp_2,\kp_4)$, $\kp_4 > \kp_4^c(\kp_2)$ where the partition function 
$Z(\kp_2,\kp_4)$ is well defined there is thus no problem with the Euclidean action being unbounded from below.
One easily shows, using the o-called Dehn-Sommerville relations, that  for a given triangulation $T$ \cite{acm,acm1}
\beq\label{j36}
 2N_4(T) \leq N_2(T).
\eeq
It is therefore easy to find regions in the coupling-constant 
space where $S_T(\kp_2,\kp_4)$ given by \rf{j32} is less than zero and 
unbounded from below when $N_2(T),N_4(T)$ goes to infinity. 
However,  in this region of coupling-constant space
$Z(\kp_2,\kp_4)$ is not well defined and it is not the region considered in EDT. In fact, in the region of large $\kp_2$, which seems 
most prone to a negative action, the limit $\kp_4 \to \kp_4^c(\kp_2)$ from above is well understood \cite{aj3} and corresponds to a 
continuum theory of fractal geometries known as random continuum trees or branched polymers. This continuum limit 
does not resemble our present universe. The fractal dimension, the  Hausdorff dimension, is two on all scales.
The same  continuum limit continues  with decreasing $\kp_2$ until one reaches a critical point $\kp_2^c$, where there is a phase 
transition, such  that for $\kp_2 < \kp_2^c$ we encounter a different kind of geometry
 when $\kp_4 \to \kp_4^c(\kp_2)$. It is  a 
``crumpled'' geometry with infinite Hausdorff dimension \cite{aj3}, where a significant fraction 
of the   four-simplices share a single 
link and the  order of the corresponding two vertices  is  very high \cite{ctkr}. 
In particular, this seems 
to be the entropically preferred type of triangulation if no curvature term is present in the action, i.e.\ $\kp_2=0$. Such  highly 
inhomogeneous   triangulations also seem unsuited to describe any theory of quantum gravity.

This leaves us with $\kp_2^c$ as the only point where one might be able to obtain an interesting theory of quantum gravity.
Potentially, this is a good scenario: at the phase transition point the typical geometries one would encounter for $N_4 \to \infty$
could be geometries with a Hausdorff dimension  between 
 $d_h = 2$ for $\kp_2 > \kp_2^c$ and the $d_h = \infty$ 
for $\kp_2 < \kp_2^c$. If the phase transition 
was  a second-order transition, this scenario could  be reasonable,
since then one might hope for a smooth transition between the two expreme limits, $d_h =2$ and $d_h= \infty$.
%
Unfortunately, the Monte Carlo simulations  show that the transition is a first-order transition and  the geometry 
at the transition point 
seems not to be a smooth interpolation between the two types of geometry \cite{bielefeld}.

This situation does not necessarily imply that an interesting continuum limit cannot
 be found, but in a Wilsonian context it implies 
that we have to use a more general action than \rf{j32}. 
Adding a suitable term, which could be a  measure term or some 
higher curvature term, in this now three-dimensional coupling constant space, 
the critical point $\kp_2^c$ will turn into a critical 
line. If the critical line ends, the endpoint would be a candidate for a 
second-order transition point. Also new phases might appear, 
and in such a more complicated landscape there might be different second-order 
(see \cite{jack} for the most recent results).

 \section{Four-dimensional CDT}
 
 Four-dimensional Causal Dynamical Triangulations (CDT) is a generalization of the simplest version of the two-dimensional 
 CDT described above. The aim is to perform the path integral over  
 geometries on a manifold $[0,1] \times \Sigma$, where $[0,1]$ denotes a   time 
 interval  and $\Sigma$ is a three-dimensional spatial manifold. 
 We discretize the time, the discrete times  labelled by 
  $t_k$. At each discretized time $t_k$ we have a spatial manifold $\Sigma$, on which we apply the EDT formalism and
  assign a piecewise linear geometry constructed by gluing together three-dimensional simplices (tetrahedra) with link lengths $a$,
 such that the topology of the three-dimensional triangulation $T^3$ (where the superscript ``3'' means that the triangulation is three-dimensional, not that it is a three-torus) matches that of the manifold $\Sigma$. Since the geodesic distances on $T^3$ are uniquely determined, so is the (piecewise linear) geometry. Given a three-dimensional 
 triangulation $T_k^3$ at time $t_k$ and a three-dimensional triangulation $T_{k+1}^3$ 
 at time $t_{k+1}$, we connect
 these by four-simplices, such that we obtain a four-dimensional triangulation with boundaries $T_k^3$ and $T_{k+1}^3$, 
 and such that the topology of this four-dimensional triangulation is $[0,1] \times \Sigma$. 
  The four-dimensional
 simplices filling out the ``slab'' between $T_k^3$ and $T_{k+1}^3$ can be of four kinds, depending on how many vertices
 the four-dimensional simplices share with a tetrahedron belonging to $T_k^3$. If the four-simplex shares four vertices with 
 $T^3_k$, i.e.\ {\it is} a tetrahedron in $T^3_k$, and thus one vertex with the triangulation $T^3_{k+1}$ we denote it a 
 $T^{(4,1)}$ simplex. If it shares three vertices with a tetrahedron in $T_k^3$, i.e.\ forms a triangle in $T^3_k$, 
 and two vertices with $T^3_{k+1}$,
 i.e.\ forms a link in $T_{k+1}^3$, we denote it a $T^{(3,2)}$ simplex. The four-simplices $T^{(2,3)}$ and $T^{(1,4)}$ are defined
 similarly. The whole construction is clearly a generalization of the construction of two-dimensional CDT, where in an analogous
 notation we would have two kinds of two-simplices, $T^{(2,1)}$ and $T^{(1,2)}$. The links of the four-dimensional simplices which are also 
 links in $T_k^3$ and $T_{k+1}^3$ are assigned a positive length $a$, while the links connecting vertices in $T_k^3$ to 
 vertices in $T^3_{k+1}$ are viewed as time-like, i.e.\ we write, in analogy with \rf{j23},
 \beq\label{j37}
 a_t^2 = - \al a^2, \qquad \al > 0.
 \eeq

 A complete triangulation of the manifold $[0,1]\times \Sigma$ is now obtained by repeating the above procedure for $k=1,2,\ldots,s$,  
 yielding a four-dimensional triangulation with spatial boundaries $T_1^3$ and $T_s^3$ and  spatial slices $T^3_k$, $1<k<s$. 
 The corresponding Regge action for such a geometry is still very simple, although slightly more complicated than \rf{j32}, since we   have introduced a parameter $\al$, which will allow us to perform a rotation of the  geometry with Lorentzian signature 
 to one with Euclidean signature. The Lorentzian action for 
 such a Lorentzian triangulation $T_{\rm lor}$, 
 expressed using the notation from \rf{j32} can be written as 
 \cite{ajl,physrep}
 \begin{eqnarray} \label{j38}
\lefteqn{S_{T_{\rm lor}} = \kp_2 \sqrt{4\alpha +1}
\Biggl[\frac{\pi}{2} N_{2}^{\rm TL}+}  \\
&& N_{4}^{(4,1)}
\Bigl( -\frac{\sqrt{3}}{\sqrt{4\alpha +1}}
{\rm arcsinh} \frac{1}{2\sqrt{2}\sqrt{3\alpha +1}}
-\frac{3 }{2} \arccos\frac{2\alpha +1}{2 (3\alpha +1)}
\Bigr)+\nonumber \\
&&N_{4}^{(3,2)}
\Biggl( \frac{\sqrt{3}}{4\sqrt{4\alpha +1}}
{\rm arcsinh}\frac{\sqrt{3}\sqrt{12 \alpha +7}}
{2 (3 \alpha+1)} - \nonumber\\
&&~~~~~~~\frac{3 }{4}
\biggl( 2\arccos\frac{-1}{2\sqrt{2}\sqrt{2\alpha +1}\sqrt{3\alpha +1}} +
\arccos\frac{4\alpha +3}{4 (2\alpha +1)}\biggr) \Biggr)\Biggr]\nonumber\\
&&-\kp_4\;\Big( N_{4}^{(4,1)}\frac{\sqrt{8\alpha +3}}{96}+
 N_{4}^{(3,2)} \frac{\sqrt{12\alpha +7}}{96}\Big).
\nonumber
\end{eqnarray}
 $N_4^{(4,1)}$ and  $N^{(3,2)}_4$ denotes the total number of 
four-simplices of types $T^{(4,1)}$ and $T^{(1,4)}$ and of types $T^{(3,2)}$ and $T^{(2,3)}$, respectively, in the triangulation $T_{\rm lor}$.
$N_2^{\rm TL}$ denotes the number of time-like triangles in the triangulation, i.e.\ triangles with one space-like link and 
two time-like links. Of course $\kp_2$ also multiplies the  number of $N_2^{\rm SL}$ of space-like triangles, but 
this number has be expressed in terms of  
the numbers $N_4^{(4,1)}$ and $N^{(3,2)}_4$ of four-simplices, by virtue of the special 
time-slicing structure present for a CDT triangulation. Finally, we have ignored a Regge boundary action term, 
coming from the two boundaries,
since in the actual computer simulations we replace the manifold $[0,1] \times \Sigma$ with the manifold
$S^1 \times \Sigma$. As we will see, the set-up of the computer simulations will be such that in most cases 
there will be no difference between choosing $[0,1]$ or $S^1$. The action is written in a way that makes it  real 
for all positive $\al$ and purely imaginary for $\al < -7/12$. Of course our starting point is a Lorentzian geometry with 
$\al > 0$, but now, like in the two-dimensional case, we can make a rotation to Euclidean geometry by performing a rotation $\al \to -\al$ in the lower
complex $\al$-plane, assuming $\al > 7/12$. One then obtains the action 
 \begin{eqnarray}\label{j39}
\lefteqn{ S_{T_{\rm eucl}}=-\kp_2\sqrt{4\tilde\alpha -1}\Biggl[\pi \Big(N_0-\chi
+\oh N_{4}^{(4,1)} +N_{4}^{(3,2)}\Big)+ }\\
&&  N_{4}^{(4,1)}\biggl( -\frac{\sqrt{3}}{\sqrt{4\tilde\alpha -1}}
\arcsin\frac{1}{2\sqrt{2}\sqrt{3\tilde\alpha -1}} +
\frac{3}{2} \arccos\frac{2\tilde\alpha -1}{6\tilde \alpha -2}\biggr)+
\nonumber\\
&&N_{4}^{(3,2)} \biggl(
+\frac{\sqrt{3}}{4\sqrt{4\tilde\alpha -1}}\arccos
\frac{6\tilde\alpha -5}{6\tilde \alpha -2}
+\frac{3}{4} \arccos\frac{4\tilde\alpha -3}{8\tilde \alpha -4}+
\nonumber\\
&&\frac{3}{2}
\arccos\frac{1}{2\sqrt{2}\sqrt{2\tilde\alpha \mi 1}
\sqrt{3\tilde\alpha \mi 1}}\biggr)
\Biggr]
 \plu\kp_4\Big( N_{4}^{(3,2)}\frac{\sqrt{12\tilde\alpha \mi 7}}{96}
\plu N_{4}^{(4,1)}\frac{\sqrt{8\tilde\alpha \mi 3}}{96}\Big).
\nonumber
\end{eqnarray}  
Analogous to  the two-dimensional case, we have  
\beq\label{j40}
S_{T_{rm lor}}(\kp_2,\kp_4,\al) \to S_{T_{\rm lor}}(\kp_2,\kp_4,-\al) = i S_{T_{\rm eucl}}(\kp_2,\kp_4,\tilde{\al}), \quad \al = \tilde{\al} > \frac{7}{12}.
\eeq  
In the same way as the constraint $\tilde{\al} > 1/4$ in the two-dimensional case was linked to the triangle inequality (and still is in \rf{j39}), the 
inequality $\tilde{\al} > 7/12$ is linked to the geometry of a $T^{(3,2)}$ simplex: for $\tilde{\al} = 7/12$ the ``time-like'' distance between 
the opposing spatial link and spatial triangle in the simplex becomes zero. In \rf{j39} we have replaced $N_2^{TL}$ by 
$N_0-\chi+\oh N_{4}^{(4,1)} +N_{4}^{(3,2)}$, where $N_0$ denotes the number of vertices in the triangulation $T$ and $\chi$
the Euler characteristic of the manifold. This relation again follows from the Dehn-Sommerville relations for four-dimensional 
CDT triangulations. One can check that for $\tilde{\al} =1$ one precisely recovers the EDT expression \rf{j32a}.

In the Monte Carlo simulations, using (generalized\footnote{We have to use slightly generalized Pachner moves to 
preserve the CDT foliation structure \cite{physrep}.})  Pachner moves to change the triangulations,  the topology of the triangulations 
is kept fixed and we can  ignore the $\chi$-term. We will do that in the following.

Let us again stress that while an expression like \rf{j39} looks 
somewhat complicated because of the $\tilde{\al}$-dependence, it is,
like \rf{j32a}, exceedingly simple, since the action of a triangulation $T$ just depends on the three global numbers 
$N_0(T)$, $N^{(4,1)}_4(T)$ and $N_4^{(3,2)}(T)$. Again the partition function for CDT quantum gravity is simply 
the generating function for the number of triangulations with given $N_0$, $N^{(4,1)}_4$ and $N_4^{(3,2)}$, with suitable indeterminates, now depending not only on $\kp_2$ and $\kp_4$, but also on $\tilde{\al}$. By redefining the coupling constants 
we can make this  simplicity explicit by writing
\beq\label{j41}
S_T(k_0,k_4,\Del) = -(k_0 + 6 \Del) N_0(T) + k_4 \big(N_4^{(4,1)}(T)+ N_4^{(3,2)}(T)\big) + \Del\, N_4^{(4,1)}(T)
\eeq
This action is still formally equal to the Regge version of the Einstein-Hilbert action for a piecewise linear manifold
constructed as described above, where the spatial links have  length $a$ and the ``time-like'' links a length $\sqrt{\tilde{\al}}\, a$.
It is still true that 
 $k_0 \propto a^2/G$, while $\Del$ is a rather complicated function of $k_0$, $k_4$ and $\tilde{\al}$ such that 
 $\Del =0$ corresponds to $\tilde{\al} = 1$. However,
 from the computer simulations to be discussed  below, it will be clear we cannot maintain such an interpretation. It is thus a more 
fruitful, Wilsonian interpretation of \rf{j41} to say that our starting point is the Regge action with
\beq\label{j42}
 \al= \tilde{\al} =1, \quad {\rm and} \quad \Del~ \mbox{is an independent coupling constant.}
 \eeq
 In this way $\Del =0$ will correspond to the Euclidean Einstein-Hilbert action \rf{j32a}, but 
 where the geometries have a time foliation
 coming from the Lorentzian geometries described above, and the new coupling constant $\Del$ is a Wilsonian enlargement 
 of the coupling-constant space from $(k_0,k_4)$ to $(k_0,k_4,\Del)$. 
 We will explore this space in the search for potentially interesting 
 phase transitions of the lattice system, which could be associated with a UV fixed point for quantum gravity\footnote{In this 
 sense the situation becomes  similar to the EDT situation, where one has to 
 enlarge the $(k_0,k_4)$ 
 coupling-constant space defined in \rf{j32a} by some new coupling constant in order to obtain an interesting result, as already mentioned.}. 
 In this context let us  mention that we can of course rewrite \rf{j41} as 
 \beq\label{j41a}
 S_T(\tilde{k}_0,k_{4,1},k_{32}) =  -\tilde{k}_0  N_0(T) + k_{41} N_4^{(4,1)}(T)+k_{32} N_4^{(3,2)}(T) 
 \eeq
 to emphasize that the action is the most general action that only depends linearly on the global number of simplices or subsimplices in a CDT triangulation\footnote{See \cite{physrep} for a classification of time- and spacelike (sub)simplices of 
 a CDT configuration, and the constraints these  numbers  satisfy. There are 10 different types of (sub)simplices and 7 constraints.}.

 To summarize, our four-dimensional CDT partition function is 
 \bea\label{j43}
\lefteqn{ Z(k_0,k_4,\Del) = \sum_T \frac{1}{C_T}\; \e^{-S_T(k_0,k_4,\Del) }}\\
 &=& \sum_{N_0,N_4^{(4,1)},N_4^{(3,2)}} 
  \e^{\big[(k_0+6\Del)N_0- k_4 \big(N_4^{(4,1)}+ N_4^{(3,2)}\big) -\Del\, N_4^{(4,1)}\Big]}  \cN(N_0,N_4^{(4,1)},N_4^{(3,2)}),\nonumber
  \eea
  where the summation is over CDT triangulations and $\cN(N_0,N_4^{(4,1)},N_4^{(3,2)})$ denotes the number of such
  triangulations with $N_0$ vertices, $N_4^{(4,1)}$  simplices of type $T^{(4,1)}$ plus type $T^{(1,4)}$, and 
  $N_4^{(3,2)}$  simplices of type $T^{(3,2)}$ plus type $T^{(2,3)}$.
  We now turn to the discussion of the phase diagram of this statistical system.
 
 \subsubsection*{Search for a UV fixed point in CDT}

\begin{figure}[t]
\centerline{\scalebox{0.8}{\rotatebox{0}{\includegraphics{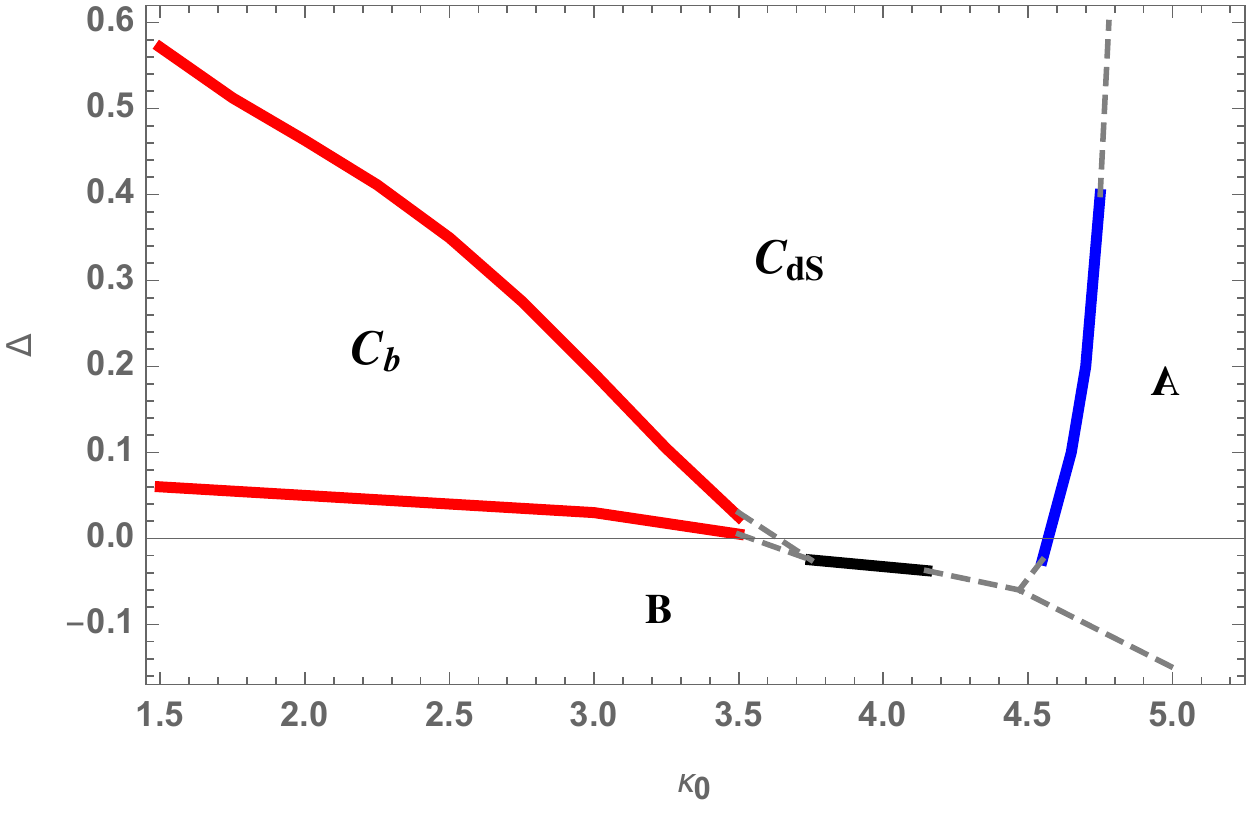}}}}
\caption{The CDT phase diagram. Phase transition between phase $C_{dS}$ and $C_b$ is 
second order when the topology of a spatial slice is $S^3$, 
as is the transition between $C_b$ and $B$. The transition between $C_{dS}$ and $A$ and the transition 
between $A$ and $B$ are first-order transitions. The transition between $C_{dS}$ and $B$ is still under investigation.}
\label{figj5}
\end{figure}

 The enlargement of the CDT coupling-constant space with the coupling constant $\Del$ leads to an amazingly complex phase 
 diagram\footnote{The phase diagram presented in the first articles \cite{ajl3,agjjl,ajjl,physrep} 
 was simpler since it missed the $C_b$ phase, discovered in \cite{signature,continuum,newhigherorder}.} 
 \cite{klitgaard} shown in Fig.\ \ref{figj5}. 
It shows the $(k_0,\Del)$ coupling constant plane. As discussed above, it is impossible to 
 keep $N_4 = N_4^{(4,1)} + N_4^{(3,2)}$ fixed in the Monte Carlo simulations. 
 However, they can be 
 conducted in such a way that {\it measurements} of the observables used to identify the phase transitions  
 are performed for a given chosen value of $N_4$. In this way  $k_4$ does not  enter actively as a coupling constant influencing 
 the observables\footnote{In many of the simulations it has been more convenient to instead keep  $N_4^{(4,1)}$ 
 fixed at the measurements.}. This is why the figure only shows the $(k_0,\Del)$ 
 coupling-constant plane. The physics
 related to the  coupling-constant $k_4$ can be recovered by performing measurements for many different values of $N_4$.
 Explicitly we have 
 \bea\label{j44}
 Z(k_0,\Del,k_4) &=& \sum_{N_4} \e^{-k_4 N_4} Z_{N_4} (k_0,\Del),\\
 Z_{N_4}(k_0,\Del) &=& \sum_{T (N_4)} \frac{1}{C_T}\;\e^{(k_0+\Del)N_0\big(T(N_4)\big) 
 - \Del \,N_4^{(4,1)}\big(T(N_4)\big)},\label{j45a}
 \eea
 where the summation is over all triangulations $T(N_4)$ with a fixed number $N_4$ 
 of four-simplices.

 Fig.\ \ref{figj5} shows the phase transition lines between the various phases, denoted $A$, $B$, $C_b$ and $C_{dS}$. 
 Here the subscript $dS$ stands for ``de Sitter'', and $b$ for ``bifurcation''. Phase $A$ and phase $B$ are most likely 
 not relevant for a four-dimensional quantum gravity theory, and the phase transitions between phases $A$ and $C_{dS}$, 
 as well as between phases  $A$ and $B$ are first-order transitions. In the case where $\Sigma$ is a three-dimensional 
 manifold with the topology of $S^3$, there is good numerical evidence that the phase transion between $B$ and $C_b$, as well
 as the transition between $C_b$ and $C_{dS}$ are second-order transitions\footnote{The order of the transition between 
 phase $C_{dS}$ and phase $B$ is most likely also a higher-order transition, but it is not entirely settled yet.}. 
 The $A$ phase can be viewed as 
 the CDT version of the branched-polymer phase of EDT and the $B$  phase as the CDT version of the crumpled phase.
 The details of these phases are discussed in Chapter 9, ``Semiclassical and Continuum Limits of Four-Dimensional CDT" in this Section of the Handbook. 
 Here we concentrate on discussing the phase transition 
 between phase $C_b$ and $C_{dS}$, since this transition 
 has a relatively transparent physical origin. 
 There is good numerical evidence that the  universe
 ``observed'' in phase $C_{dS}$ can be considered as homogeneous and isotropic in the spatial directions, while the 
 homogeneity is broken in phase $C_b$. This symmetry breaking appears to happen smoothly (in accordance with the 
 higher-order nature of the transition) \cite{signature,klitgaard}, such that regions of inhomogeneity become more and more pronounced the 
 deeper we move into phase $C_b$, starting from the border between phase $C_b$ and phase $C_{dS}$. It is tempting 
 to conjecture that the seeds of inhomogeneity of geometry at the phase transition line could act as seeds for matter 
 inhomogeneity, in case  matter was added to the model, 
 and maybe in this way be important for the first galaxy 
 formations. In particular this would present an intriguing  scenario if the phase transition line could be associated with 
 a UV fixed point. First of all, as already mentioned, 
 a UV fixed point in a theory of gravity is central to  the 
 asymptotic safety scenario, and  finding it in our lattice approach is central to the idea that one can use 
 the lattice theory as a non-perturbative definition of  quantum gravity. 
 Secondly, it is also often assumed that the ``origin'' of the Universe 
 is associated with the theory of gravity at short distances (some kind of Big Bang scenario), and this should then naturally
 relate to physics close to the UV fixed point. Inhomogeneity as part of  this UV physics could then be important for the formation
 of  structure in a universe with matter.

 Therefore, confronted with a second-order phase transition
 line, the phase transition line between the $C_b$ and $C_{dS}$ phases, the obvious question of interest is
 whether or not there is a non-perturbative UV fixed point associated with this line. 
 
 \subsubsection*{Search for a UV fixed point in a $\phi^4$ theory}
 
 In order to 
 address this question, let us step back and briefly recall how it has been addressed in ordinary $\phi^4$ scalar 
 field theory in four-dimensional flat (Euclidean) spacetime\footnote{Since the Higgs field $\phi$ in the Standard Model 
 is governed by a $\phi^4$ field theory (embedded in a larger theory), the existence or non-existence of such a UV fixed 
 point is actually important for Standard Model.}. Consider the scalar $\phi^4$ theory defined on a four-dimensional 
 hyper-cubic lattice with lattice spacing $a$. We denote  the integer lattice coordinates of the vertices by
 $n =(n^1,\ldots,n^4)$ and the spacetime coordinates of these lattice points by $x_n=a n$.
A lattice scalar field $\phi$ takes values on the lattice vertices and we use the notation $\phi(n)$ or $\phi(x_n)$.
The action is 
\beq\label{s2-3}
S[\phi,\mu,\lam;a] = \sum_{n} a^4 \Big( \oh\sum_{i=1} ^4\frac{(\phi(n\!+\!\hat{i}\,) \!-\!\phi(n))^2}{a^2} + 
\oh \frac{\mu}{ a^2}\,\phi^2(n) +\frac{1}{4!}\lam \phi^4(n)\Big),
\eeq
where $\hat{i}$ denotes  the unit vector in direction $i$.

The theory has two dimensionless  lattice coupling constants $\mu$ and $\lam$. In this coupling-constant 
space there is a phase transition line  between a symmetric phase 
where $\la \phi(n) \ra =0$ and a symmetry-broken phase where $\la \phi(n) \ra \neq 0$. The symmetry broken is 
$\phi (n) \to - \phi (n)$. This transition line is a second-order phase transition line, and the correlation length between
the fields at different lattice points diverges when one approaches the transition line. The question is whether  
 this phase transition line be used to define a non-perturbative UV fixed point for 
 the  $\phi^4$  quantum field theory. 
 The tentative continuum quantum field theory is defined 
 by its two renormalized continuum coupling constants
 $m_R$ and $\lam_R$, the continuum mass and the continuum $\phi^4$ coupling constant. 
 They can be extracted from the two-point correlator and the four-point correlator.  The lattice two-point 
 function is characterized by a lattice correlation length $\xi$. We can write
 \beq\label{j45}
 \xi(\mu,\lam)  = \lim_{|n-n'| \to \infty} \frac{ - \log \big\la (\phi(n)-\la \phi\ra) (\phi(n')-\la \phi \ra)\big\ra}{ | n-n'|}, 
 \qquad m_R = \frac{1}{a\, \xi}.
 \eeq
 We assume for simplicity that the coupling constants are chosen such 
 that we are in the symmetric phase, where $\la \phi \ra =0$.
 Let us not discuss in detail how to define $\lam_R$ (for details see for instance \cite{munster}), but only state that insisting that $\lam_R$ is constant defines a path 
 $(\mu(\xi),\lam(\xi))$ in the lattice coupling-constant space  $(\mu,\lam)$. For each point on this path
 we can calculate a correlation length $\xi$ using \rf{j45}, and we use these $\xi$ as a parametrization of the path. If the 
 path meets the second-order transition line at a point  $(\mu^c, \lam^c)$, it implies that $\xi \to \infty$ at this point.
 This point can  serve as a UV fixed point, since we now demand that $m_R$ is constant along the path, i.e.\ the lattice 
 spacing $a$ becomes a function of $\xi$ via \rf{j45},
 \beq\label{j46a}
 a(\xi) = \frac{1}{m_R \xi}, \quad {\rm i.e.} \quad a(\xi) \to 0 \quad {\rm for} \quad \xi \to \infty.
 \eeq
 We conclude that  {\it by  following a path in the bare, 
 dimensionless coupling-constant space, where 
 continuum observables are kept fixed, one is led to a UV fixed point, provided it exists.} If the UV point does not exist, the path
 will be such that $\xi$ never reaches infinity, no matter where we start in the bare coupling constant space. According to 
 \rf{j46a} this implies that we cannot remove the UV cut-off $a$.
  
 The approach to  the UV fixed point is governed by the $\beta$-function\footnote{
The $\beta$-function is a function of $\lam$ and $\mu$, but close to the fixed point
one can ignore the $\mu$-dependence.}, 
which relates the change in $\lam$ to the change in $a(\xi) = 1/(m_R \xi)$ as we move along
the trajectory of constant $m_R,\lam_R$,
\beq\label{s2-1}
-a \frac{\d \lam}{\d a}\Big|_{m_R,\lam_R} = \xi \frac{\d \lam}{\d \xi}\Big|_{m_R,\lam_R}= \beta(\lam) .
\eeq
Since $\lam(\xi)$ 
stops changing when $\xi \to \infty$, we have $\beta(\lam^c) =0$, and expanding the $\beta$-function 
to first order one finds
\beq\label{s2-2}
\lam(\xi) = \lam^c + const. \;\xi^{\beta'(\lam^c)},\qquad
\beta'(\lam)= \frac{d \beta}{d \lam}.
\eeq
It follows from \rf{s2-2} that  $\beta'(\lam^c)< 0$ at  a UV fixed point.

The correlation length $\xi$ clearly plays a major role in the above scenario. It will be convenient 
to replace it with a finite lattice volume by using so-called finite-size scaling. Assume 
we have a finite hypercubic lattice. The volume is then $V= N a^4$, where $N$ is the number of hypercubes. 
We keep the ratio between the linear size of the lattice and the correlation length fixed, 
\beq\label{s2-6}
\frac{\xi}{N^{1/4}}= \frac{ 1}{(a(\xi)m_R) N^{1/4}} = \frac{1}{m_R V^{1/4}}.
\eeq 
Thus, if we  are moving along a trajectory with constant $m_R$ and  $\lam_R$ in the 
bare $(\mu,\lam)$-coupling constant plane and change $N$
according to \rf{s2-6}, the finite  continuum volume stays fixed. Assuming that there is a
UV fixed point, such that $a(\xi) \to 0$, we see that $N$ can go to infinity 
even if $V$ stays finite, and 
that the correlation length $\xi$ in \rf{s2-2} can be substituted by a dependence on the linear size $N^{1/4}$ in lattice units of the spacetime, leading to
\beq\label{2.2}
\lam(N) = \lam^c + const. \;N^{\beta'(\lam^c)/4}.  
\eeq
As we saw above, the  absence of a UV fixed point could be deduced by the absence of a divergent 
correlation length along a trajectory of constant physics in the $(\mu,\lam)$-plane. 
In the finite-size scaling scenario this is restated as 
$N$ not going to infinity along any such curve of constant physics.  In the case of a $\phi^4$ theory in four-dimensional
spacetime this is exactly what happens, and the conclusion is that there is no UV fixed point in the theory \cite{luscher,luscher1}. The 
second-order transition line of the theory is related to the IR limit of the theory where $\lam_R =0$.

\subsubsection*{Finite-size scaling analysis in  CDT}

\begin{figure}[t]
\centerline{\scalebox{0.9}{\rotatebox{0}{\includegraphics{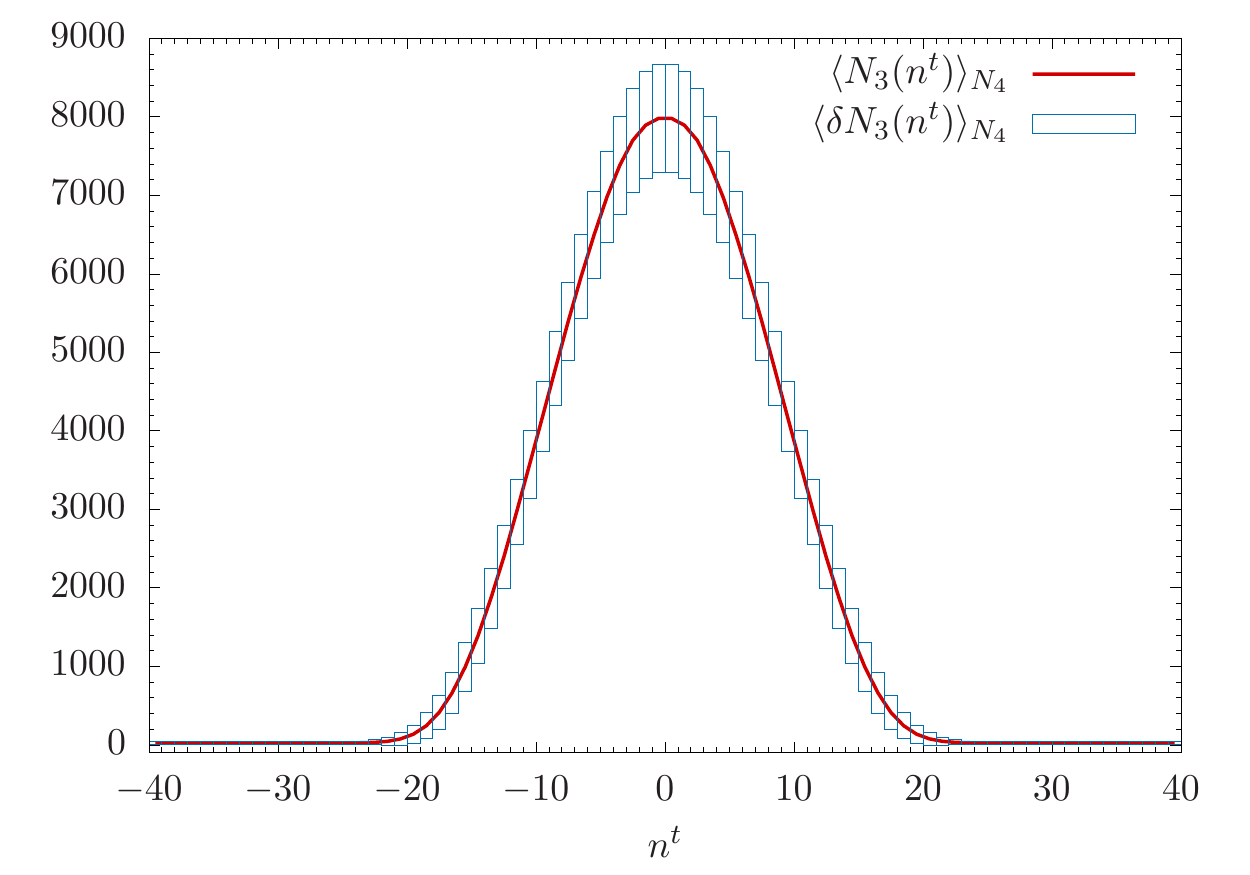}}}}
\caption{The average spatial volume $\langle N_3(t_k)\rangle_{N_4}$ as a result of 
MC measurements for  $N_4=362.000$. 
The  best fit of the form \rf{j52} yields a curve which cannot be distinguished from the 
measured average at the given plot resolution.
The bars indicate the average size of quantum fluctuations $\del N_3(t_k)$.}
\label{figj6}
\end{figure}

We now want to apply the above formalism to CDT \cite{renorm1,renorm2} 
and in addition  take advantage of the time-slice structure present in 
CDT. In fact, one does precisely that in  Monte Carlo simulations on an ordinary hypercubic lattice. Instead of the point-point 
correlator $\la \phi(n) \phi(n')\ra$ used in \rf{j45}, one averages the positions $n$ and $n'$ over positions $n({t_k})$ belonging
to a hyperplane located at ``time'' $t_k= k a$  and positions $n'({t'_{k'}})$  
located at   ``time''  $t'= k' a$, where the two hyperplanes are separated by
a lattice distance $d = |k-k'|$. This reduces the fluctuations of the measured correlator and it also has the advantage that 
power corrections to the exponential fall-off of the correlator are absent, making it easier to determine the correlation length
(see again \cite{munster} for details).
Thus one replaces \rf{j45} by 
\beq\label{j46}
\left\la \sum_{n({t_k})} \phi(n({t_k})) \sum_{n'(t_{k'}')} \phi(n'({t'_{k'}}))\right\ra = const. \; \e^{-|k-k'|/\xi}.
\eeq
In our CDT theory of pure geometry we do not have a field $\phi(n)$ at our disposal, but as in the two-dimensional case 
we can use the ``unit'' field $1(n)$ which assigns the value 1 to each four-simplex. 
Each three-simplex $T^3$ in the three-dimensional triangulation $T^3_k$ corresponding 
to time $t_k$ belongs to two  four-simplices $T^{(4,1)}(T^3)$ and $T^{(1,4)}(T^3)$. 
We can then write
\beq\label{j47}
\sum_{n(t_k)} \phi(n(t_k)) \to \oh \sum_{T^3 \in T^3_k} \Big( 1\big(T^{(4,1)}(T^3)\big) + 1\big(T^{(1,4)}(T^3)\big) \Big)
= N_3 (t_k),
\eeq
where $N_3(t_k)$ is the number of three-simplices in the spatial slice at time $t_k$.
Since $\la N_3(t_k)\ra > 0$ we have a situation like in the 
broken phase of a $\phi^4$ theory: connected correlators have to be expanded 
around $\la \phi \ra \neq 0$. Here one has to  expand around $\la N_3(t_k)\ra > 0$
(see eq.\ \rf{j50} below). Moreover, it turns out
that  $\la N_3(t_k)\ra$ will be a highly non-trivial function of $t_k$, see Fig.\ \ref{figj6},
once the translational invariance of the action in $t$ is dealt with in the proper way. 
In the symmetric phase of the $\phi^4$ theory all correlators with an odd number of fields 
will be zero, and  the two- and the four-point correlators are needed in order to study the renormalization flow of $\mu$ and 
$\lam$. Here, in CDT, we can extract information of the flow of $k_0$ and $\Del$ by considering only one- and two-point 
correlators because the one-point function $\la N_3(t_k)\ra$  
has a non-trivial dependence on $t_k$\footnote{The non-trivial dependence 
on $t_k$ is shown in Fig.\ \ref{figj6}, and it only appears after the zero mode corresponding 
to translational invariance in $t$ has been eliminated. The condition $H(0) =1$ then
refers to the time $t_0=0$ that is chosen as the maximum of the ``blob" $\la N_3(t_k)\ra$.}
\bea\label{j48}
\la N_3(t_k) \ra_{N_4}& \propto & N_4^{3/4}  \;H\left(\frac{k}{N_4^{1/4}}\right), \quad H(0) =1\\
\big\la  N_3(t_k) N_3(t'_{k'}) \big\ra^c_{N_4} &\propto & N_4 \;F\left(\frac{k}{N_4^{1/4}}, \frac{{k'}}{N_4^{1/4}}\right),\label{j49}
\eea
where the connected correlator, as in \rf{j45},  is defined by 
\beq\label{j50}
\big\la  N_3(t_k) N_3(t'_{k'})  \big\ra^c_{N_4} = \big\la (N_3(t_k)- \la N_3(t_k) \ra)( (N_3(t'_{k'})- \la N_3(t'_{k'}) \ra) \big\ra_{N_4}.
\eeq
 In particular we have 
 \beq\label{j51}
 \la \del N_3(t_k) \ra_{N_4} := \sqrt{\la  N_3^2(t_k) \ra^c_{N_4}}\propto \sqrt{N_4} \;F\left(\frac{k}{N_4^{1/4}}, \frac{k}{N_4^{1/4}}\right), \quad F(0,0) =1 .
 \eeq
 Monte Carlo simulations confirm the functional form alluded to in \rf{j48} and \rf{j49} when we are in phase $C_{dS}$. 
 In particular we have measured with high precision \cite{agjl,agjl1}
 \beq\label{j52}
 H\left(\frac{k}{N_4^{1/4}}\right) \propto \cos^3 \left( \frac{k}{\om N_4^{1/4}}\right)
 \eeq
 where $\om$ depends on $k_0$ and $\Del$. This functional form is the reason why we call phase $C_{dS}$ the 
 ``de Sitter'' phase. It is the functional form that $N_3(t_k)$ would have for a four-sphere where $t_k$ denotes the 
 geodesic distance from the three-equator. It is  valid for 
 \beq\label{j53}
 -\frac{\pi}{2} \om N_4^{1/4} < k < \frac{\pi}{2} \om N_4^{1/4},
 \eeq
 where we for convenience have chosen to locate 
 the maximum of $\la N_3(t_k) \ra_{N_4}$  at $k=0$.
 In the computer simulations which resulted in this distribution, the lattice time extension was chosen larger than 
 $\pi \om N_4^{1/4}$. Outside the region \rf{j53} one finds $N_3(t_k) \approx 0$ which is
 of the order of the cut-off, i.e.\ 
 the smallest $S^3$ one can create by gluing together 5 tetrahedra). This is why it does not 
 matter whether we choose the time direction to correspond to $S^1$ or to $[0,1]$. Fig.\ \ref{figj6} shows the 
 measured three-volume profile \rf{j48},  as well as the theoretical curve \rf{j52}, and finally the fluctuations \rf{j51} around the 
 measured three-volume profile.
 
 Eqs.\ \rf{j48} and \rf{j51} allow us  in principle to follow  a path of constant continuum physics in the $(k_0,\Del)$ coupling-constant 
 space, which might lead to a UV fixed point,  
 in the spirit of the finite-size scaling discussion for the $\phi^4$ theory. 
 We define the continuum three-volume of a time slice as 
 \beq\label{j54}
 V_3 (t_k) = \frac{\sqrt{2}}{12} a^3 N_3(t_k),  \qquad t_k = k \,a.
 \eeq
 First, for fixed $(k_0,\Del)$, taking $N_4 \to \infty$, we see that 
 \beq\label{j55}
 \left.\frac{ \del V_3(t_k)}{V_3(t_k)}\right |_{N_4} = 
 \left.\frac{ \la \del N_3(t_k) \ra}{\la N_3(t_k) \ra}\right |_{N_4}  \propto \frac{1}{N_4^{1/4}} \to 0\quad {\rm for} \quad N_4 \to \infty.
 \eeq
 The simplest interpretation of this result is that for fixed $(k_0,\Del)$ we should view the
 lattice spacing $a$ in \rf{j54} as constant. Then
 $N_4 \to \infty$ implies that  $V_3 \to \infty$ and  for large continuum $V_3(t)$ 
 one expects that the 
 the fluctuations will be small relative to $V_3(t)$. However, in 
 the spirit of finite-size scaling, we are interested in a limit where $V_3(t)$ stays finite when $N_4 \to \infty$. This is clearly a limit 
 where  also the fluctuations around $V_3(t)$ will stay finite, since they then represent the ``real'' continuum fluctuations around $V_3(t)$, and they should be independent of $N_4$
 for sufficient large $N_4$. We therefore require
 \beq\label{j56}
 \frac{ \del V_3(t)}{V_3(t)} = \frac{ \la \del N_3(t) \ra}{\la N_3(t)\ra} = const. \quad {\rm for~fixed} \quad
 t = k \,a \propto k \, N_4^{-1/4}. 
 \eeq
 According to \rf{j55}, we can only obtain this by changing $(k_0,\Del)$ when we change $N_4$. We thus have
 precisely the picture advocated in the $\phi^4$ case:
  {\it  the requirement of constant continuum physics leads to 
 a path in the bare, dimensionless lattice coupling-constant space when we increase $N_4$.
 If this path continues to $N_4 \to \infty$, then
 according to \rf{j54} the lattice spacing $a \to 0$ and we might 
 reach a UV fixed point $(k_0^c,\Del^c)$.}

 In \cite{renorm1} it was attempted to follow this program, but it was before the 
 $C_b$-$C_{dS}$ phase transition line 
 was discovered and the $N_4$ used 
 might have been too small, so the question about the existence of a UV fixed point associated with the 
 $C_b$-$C_{dS}$ phase transition line is still open.
 
 A different approach, in some sense closer to the renormalization group approach, is to use the Monte Carlo simulation
 data to construct an effective action for the three-volume $N_3(t_k)$. The 
 corresponding effective action can be viewed as a kind of  {\it minisuperspace action}. However,  the action is 
 (numerically) derived from the full quantum theory, so the geometric 
 degrees of freedom different from $N_3(t)$ have not been ignored, but rather (numerically) integrated out. 
 Chapter 9, ``Semiclassical and Continuum Limits of CDT" in this Section of the Handbook will discuss this in detail. So far, it has not been possible to identify in an unambiguous way 
 a UV fixed point, but the formalism for doing so now exists, as explained above, and 
 future computer simulations can hopefully clarify if the fixed point exists or not.

 \section*{Future perspectives}
 
 In discussing four-dimensional CDT we have focused on how one can in principle 
 locate a UV fixed point. If it exists and if it is non-trivial, it would be a strong indication
 that there exists a non-perturbative, unitary quantum field theory of Lorentzian geometries
 at all length scales. It could be the quantum theory of GR. Of course one would 
 have to provide convincing arguments in favor of such an interpretation. The CDT 
 theory is most likely unitary when rotated back from Euclidean spacetime to 
 Lorentzian spacetime, since one can show that the Euclidean rotated version
 of CDT is reflection positive, a property that  usually ensures that when one 
 rotates back to Lorentzian signature, one obtains a unitary theory. We expect 
 a quantum theory of GR to be unitary, so a putative continuum  4d CDT theory
 passes that test. It also makes in unlikely that the continuum limit of 4d CDT
 should be some generic $R^2$ version of GR, since these theories typically
 will be non-unitary theories.  
 Another test is that classical GR should emerge 
 from the quantum effective action in the limit where $\hbar \to 0$. Such a test 
 could be performed if one could construct the effective action of the 
 quantum theory. As mentioned the effective action has been constructed for 
 the three-volume $V_3(t)$ and it is closely related to a GR  minisuperspace
 action. However, the real test would be to construct the the full quantum
 effective action from the MC data and take the $\hbar \to 0$ limit, 
 and we do not yet know how to do that.
 Maybe the simplest way to relate the lattice theory associated with a UV fixed 
 point to  continuum gravity theories is    
 to determine the critical exponents related to the lattice fixed point 
 and compare with similar analytic renormalization group calculations. 
 In principle the functional  renormalization group approach should provide us with 
  unique critical exponents if a  UV quantum gravity fixed point exists. 
 In practice the calculations in both approaches will have error-bars and the comparison
 might not be easy. 
 
There is a number of conceptional issues in a theory of 
quantum gravity. We tried to  illustrate these in 
the case of solvable two-dimensional gravity models.
 How do we talk about distances in a theory of quantum gravity 
 where  in the path integral we integrate over the 
 geometries  that determine distances? Does it make sense to talk about arbitrarily small distances, much smaller 
 than the quantum fluctuations of geometries? The solvable models of two-dimensional gravity encourage us to believe
 that it can make sense, and that we are not forced to endorse 
 the common statement that "the concept of geometry has to break 
 down at short distances". Of course the situation could be different 
 in four dimensions and there is little hope that we can solve 
 the four-dimensional theories analytically. However, 
 the Monte Carlo simulations of the four-dimensional 
 quantum gravity models might teach us how we {\it should} think about geometry at the shortest distances. In the 
 two-dimensional models there have been very fruitful interplays between pure theory and Monte Carlo simulations of the 
 models. What can relatively easily be measured in the Monte Carlo simulations are the fractal dimensions of the spacetime
 geometries, both the so-called Hausdorff dimension and spectral dimension
 \footnote{This is discussed Chapter 3, ``Spectral Observables and Gauge Field Couplings
 in Causal Dynamical Triangulations'' in this Section of the Handbook.}. 
 Let us just mention that the measurement of the spectral 
 dimension in 4d CDT resulted in the surprising result 
 that the dimension seems to be {\it scale-dependent} \cite{spectraldim}. Inspired by this,  similar results have been obtained
  analytically 
 in a number of quantum gravity models. This is an example of a fruitful interplay between numerical studies and 
 analytic calculations also in higher than two dimensions.   
 
 Presently we do not know for certain if there exists a UV fixed point that will allow us to define an ``ordinary'' quantum field 
 theory of quantum gravity at all scales. However, 
  the lattice efforts will not be wasted even if it should turn out that such a fixed point does 
 not exist. First of all there exists most likely in ordinary continuum gravity an effective quantum field theory 
 up to energies of the order of the Planck energy. There is no conceptual problems 
 also using such an effective theory in a cosmological context. It will be a quantum field theory with a cut-off. The lattice theories,
 both four-dimensional  EDT and CDT will be such theories, where the lattice spacing acts as the cut-off. Such cut-off theories
 can still provide us with  a lot of non-perturbative  information about (the effective theory of) quantum gravity, since
 non-perturbative information is not necessarily linked to short-distance phenomena. In fact, an example of this is provided 
 by one of the first  cases where non-perturbative topological aspects entered into quantum field theory, namely in the 
 three-dimensional Georgi-Glashow model. This model is invariant under local $SO(3)$ gauge transformations. It has a non-Abelian gauge field  and a three-component scalar field  that transforms as a vector under the action of the $SO(3)$ gauge group. 
 When rotated to Euclidean space-time the model has classical monopole solutions, that act as instantons. The Higgs potential
 is such that the $SO(3)$ symmetry is broken down to a $U(1)$ symmetry with an associated Abelian gauge field, 
 while the two other gauge field components combine  to a massive charged, 
 so-called $W$ particle.  
 The monopoles will result in confinement of particles with $U(1)$ charge \cite{polyakov}. 
 Amazingly, a pure $U(1)$ lattice gauge theory will produce exactly the same non-perturbative 
 physics because it has lattice monopoles \cite{polyakov1}. 
 These monopoles have a mass which diverges as $1/a$, the lattice spacing,
 while the mass of the monopoles in the Georgi-Glashow model is proportional to $m_w/e^2$, where $m_w$ is the mass 
 of the $W$ particle and $e$ the charge associated to the $U(1)$ symmetry. The lattice physics is therefore identical to the 
 physics of the Georgi-Glashow model for distances larger than the lattice spacing $a$, as long as $a \sim e^2/m_w$. 
 However, we cannot take $a < e^2/m_w$ and still capture the physics of the Georgi-Glashow models. 
 When $ a\to 0$ the lattice monopoles will be infinitely heavy and decouple,
  and we will just obtain a theory with a free photon. If we have a situation in quantum gravity,  where there is no 
  UV fixed point, there might be new degrees of freedom for distances shorter than the  Planck length, and
  we will  not be able to represent them correctly when our lattice spacing $a$ is less than the Planck length. But 
  physics at scales larger than the Planck length, including some non-perturbative physics caused by these unknown degrees 
  of freedom, could still be correctly described by our lattice models. In the case of CDT we have, as mentioned above,  constructed an 
  effective minisuperspace action, which may 
  describe physics well  all the way down to  the Planck length and 
  allow us to study universes which are not much larger than the Planck length, and discover
  corrections to the simplest minisuperspace models. In addition, 
  we can also study the non-perturbative interaction
  between matter fields and gravity in a full quantum context\footnote{The non-trivial 
  interaction between geometry and matter is described in Chapter 8, ``Scalar Fields in 
  Four-Dimensional CDT", of this Section of the Handbook}. 
  But hopefully there 
  {\it is} a UV fixed point. Then such studies will not be confined to distances larger than the Planck length.

\end{document}